\documentclass[letterpaper,aps,reprint,prd,nofootinbib,showpacs,noshowkeys,longbibliography]{revtex4-1}
\usepackage{hyperref} 
\usepackage{amsmath}
\usepackage{amsfonts}
\usepackage{amsthm}
\usepackage{amssymb}
\usepackage{cancel}
\usepackage{graphicx}
\usepackage{feynmf}
\usepackage{bbding}
\usepackage{slashed}
\usepackage[sort&compress]{natbib}

\newcommand*\Temp{\tilde{T}} 
\newcommand*\Erg{\tilde{\mathcal{E}}} 
\newcommand*\Action{\tilde{S}} 
\newcommand*\Entropy{\tilde{\Sigma}} 

\begin{document}

\title{Thermal Schwinger pair production at arbitrary coupling}
\date{June 12, 2017}
\author{Oliver Gould}
\email{o.gould13@imperial.ac.uk}
\author{Arttu Rajantie}
\email{a.rajantie@imperial.ac.uk}
\affiliation{Department of Physics, Imperial College London, SW7 2AZ, UK}



\begin{abstract}
We calculate the rate of thermal Schwinger pair production at arbitrary coupling in weak external fields. Our calculations are valid independently of many properties of the charged particles produced, in particular their spin and whether they are electric or magnetic. Using the worldline formalism, we calculate the logarithm of the rate to leading order in the weak external field and to all orders in virtual photon exchange, taking us beyond the perturbative expansion about the leading order, weak coupling result.
\end{abstract}

\maketitle

\section{Introduction}

In the presence of an electric field, empty space is unstable to the production of electron-positron pairs, called Schwinger pair production \cite{schwinger1951gauge}. The usual perturbative vacuum is not the true vacuum, the lowest energy state, and hence it decays. At finite temperature, the energy available from the thermal bath enhances the rate of decay.

For weak coupling much has been done to generalise Schwinger's original result, including the effect of temporal and spatial variation in the external field \cite{brown1964interaction,nikishov1964quantum,brezin1970pair,nikishov1970barrier,wang1988finite,kim2002schwinger,gies2005pair,dunne2005worldline}; the presence of an additional high energy photon or other particle \cite{selivanov1985destruction,popov2001schwinger,dunne2009catalysis,monin2010photon,monin2010semiclassical}; a finite temperature \cite{cox1984finite,selivanov1986tunneling,elmfors1994electromagnetic,elmfors1998thermally,kim2010nonperturbative,gavrilov2008one,monin2008semiclassical,king2012pair,brown2015schwinger,medina2015schwinger}; higher loops \cite{ritus1975lagrange,ritus1977connection,lebedev1984virial,ritus1998effective,gies2000qed} and back reaction \cite{smolyansky1997dynamical,kluger1998quantum,alkofer2001pair,hebenstreit2013real,hebenstreit2013simulating}. However, at stronger coupling, where perturbation theory breaks down, much less is known.

In this paper we calculate the rate of Schwinger pair production from a thermal bath, making no assumptions about the strength of the coupling. We do though restrict ourselves to weak external fields. Our results are argued to be valid for the full range from zero to infinite coupling \cite{affleck1981pair,lebedev1984virial} (see, however, \cite{dunne2004multiloop,huet2010euler,huet2017asymptotic}).

There are many applications of this calculation but our interest stems from the wish to better understand the pair production of magnetic monopoles \cite{dirac1931quantised,thooft1974magnetic,polyakov1974particle}. Theoretical understanding of this is poor and there is a pressing need for concrete calculations of rates and cross sections for ongoing experimental searches such as MoEDAL \cite{moedal2016search} at CERN. The Dirac quantisation condition implies that if magnetic monopoles exist, they are necessarily strongly coupled. Their charge, $g$, must satisfy $g=2\pi n/e\approx 20.7 n$, where $n\in \mathbb{Z}$ and $e$ is the charge of the positron. As a result of this, calculations are difficult and cross sections for their pair production are extremely poorly understood.

In the collisions of \emph{elementary particles} it has been argued that the pair production of 't Hooft-Polyakov magnetic monopoles \cite{thooft1974magnetic,polyakov1974particle} is exponentially suppressed by \cite{drukier1982monopole}
\begin{equation}
\mathrm{e}^{-16\pi/e^2}\approx 10^{-238}, \label{eq:suppression}
\end{equation}
even at arbitrarily high energies. The suppression can be seen as due to its large, coherent structure and the small overlap of this state with the initial, perturbative state. An analogous exponential suppression has been explicitly demonstrated for soliton pair production in a particular scalar theory \cite{demidov2011soliton,demidov2015high,demidov2015semiclassical}. Note though that this argument does not apply to pointlike, Dirac monopoles. For them, the lack of any small parameter has meant that there have been no estimates of pair production cross sections which have been derived from first principles. Consequently accelerator searches for magnetic monopoles can only yield upper bounds on the production cross section \cite{rajantie2016search}, they cannot constrain the mass, spin or charge of magnetic monopoles.

In this paper, we consider a different process, relevant to physical situations where there are strong magnetic fields and high temperatures. In such situations pair production proceeds via the magnetic dual of Schwinger pair production, as first considered at zero temperature by Affleck, Manton and Alvarez \cite{affleck1981monopole,affleck1981pair}. For this process the arguments of Ref. \cite{drukier1982monopole} do not apply. Further we can calculate the rate of pair production from first principles, the result being valid for both Dirac and 't Hooft-Polyakov monopoles (see appendix \ref{appendix:extended}).

In particular, strong magnetic fields and high temperatures arise in heavy ion collisions \cite{alice2013measurement,skokov2009estimate}, though not in elementary particle collisions. The difference is crucial and our results suggest that, for sufficiently light magnetic monopoles, pair production in heavy ion collisions is not exponentially suppressed as in Eq. \eqref{eq:suppression}. Like the case of $(B+L)$ violation \cite{thooft1976symmetry,thooft1976computation,klinkhamer1984saddle,kuzmin1985anomalous,arnold1987sphalerons,arnold1987sphaleron}, we suggest that this is because, in the initial thermal state, the energy is spread across many degrees of freedom.

We can study magnetic monopole Schwinger pair production via its electromagnetic dual as we do not consider electric and magnetic charges simultaneously. In this case the duality amounts simply to a relabelling of electric degrees of freedom and charges as magnetic. We will, in the bulk of the paper, refer to pair production of particles with charge $g$ in an external field $E$, whether electric or magnetic. The mass of the charged particles is denoted by $m$. As our calculation reduces to a semiclassical one, we only rely on the classical electromagnetic duality.

Our calculation is also of relevance to Schwinger pair production of atomic nuclei, especially those with charges $Ze\gtrsim 1$ ($Z\gtrsim 3$) where the usual weak coupling approaches break down, and to pair production of quarks in QCD, in the Abelian dominance approximation \cite{casher1978chromoelectric,casher1979multiparticle,bialas1987oscillations,ganguly1995thermal,florkowski2003schwinger}. Further it gives an all-orders correction to the known weak coupling results. This will be of interest to current and future experimental studies of Schwinger pair production (see for example \cite{ringwald2003boiling} for a discussion), as well as to multi-loop and asymptotic analyses of the QED perturbation series \cite{dunne2004multiloop,huet2010euler,huet2017asymptotic}.

In Sec. \ref{sec:general} we set up our calculation using the worldline formalism, deriving an expression for the thermal Schwinger rate at arbitrary coupling. In Sec. \ref{sec:approximation} we explain a key approximation that we make, the dilute instanton gas approximation. In Sec. \ref{sec:instantons} we derive analytic results in various limits and in Sec. \ref{sec:numerical_results} we extend beyond these limits via numerical calculations. We also discuss the general form of the rate in terms of a phase diagram. In Sec. \ref{sec:conclusions} we conclude and suggest further work.

Throughout we use natural units, $c=\hbar=k_B=1$.

\section{Worldline expression\label{sec:general}} 

\subsection{Zero temperature rate}

Physically, we consider at an initial time a state, such as a thermal state. We choose the state such that, in the absence of an external field, there are no net production or annihilation rates. If we then adiabatically turn on an external field, the initial state becomes unstable to a net production of charged particles. We wish to calculate this rate.

We denote by $|\Omega\rangle$ the zero temperature state in the absence of the external field, the so called false vacuum. The probability of the decay of this state is given by
\begin{equation}
 P=1-|\langle \Omega | \hat{S} | \Omega \rangle|^2=1-\mathrm{e}^{-\mathcal{V}\Gamma},
\end{equation}
where $\hat{S}$ is the S-matrix including the external field. As both the false vacuum state and the external field are homogeneous, the quantity of interest is the probability per unit spacetime volume, or the rate per unit volume, $\Gamma$. The rate of pair production is given by twice the imaginary part of the energy density of the initial state.
\begin{align}
 \Gamma &= \frac{2}{\mathcal{V}} \mathrm{Im}( -i \log\langle \Omega | \hat{S} | \Omega \rangle) \nonumber \\
 &= 2 \mathrm{Im}(\mathcal{E}), \label{eq:minkowskirate}
\end{align}
where $\mathcal{V}$ is the volume of spacetime and $\mathcal{E}$ is the energy density of the false vacuum. Note that for this to make sense we should do the calculation in a finite volume and take the volume to infinity at the end.

We can analytically continue Eq. \eqref{eq:minkowskirate} to Euclidean time, as the energy density can equally well be calculated in Euclidean time. It then becomes \cite{stone1975lifetime}
\begin{equation}
\Gamma = \frac{2}{\mathcal{V}} \mathrm{Im}( -\log\langle \Omega | \hat{S}_E | \Omega \rangle),  \label{eq:rate_imag} \\ 
\end{equation}
where $\hat{S}_E$ refers to the ``S-matrix'' corresponding to Euclidean time evolution. The generalisation of this result to non-zero temperatures can then easily be made. Using the Matsubara formalism, finite temperatures simply correspond to finite Euclidean time extents and periodic boundary conditions \cite{bloch1932zur}.

We consider quantum electrodynamics (QED) and scalar quantum electrodynamics (SQED) in 4D flat spacetime. In the worldline formalism these two theories are related, the only difference being the presence of a spin factor in the QED worldline path integral. In Appendix \ref{appendix:qedsqed} we show that the spin factor does not turn up in the leading term for weak external fields. As we will make this approximation below, we restrict our attention to SQED, the final results also being valid for QED. SQED is the model of a photon, $A_\mu$, interacting with a massive charged scalar particle, $\phi$, with charge $g$. The introduction of the external field, $A_\mu^{\mathrm{ext}}$, is achieved by shifting the gauge field, $A_\mu\to A_\mu+A_\mu^{\mathrm{ext}}$, in the covariant derivative of $\phi$. The Euclidean Lagrangian is then
\begin{equation}
 \mathcal{L}_{\mathrm{SQED}}:=\frac{1}{4}F^{\mu \nu}F_{\mu \nu} + D_\mu \phi ( D^\mu \phi)^* + m^2\phi\phi^*, \label{eq:sqedlagrangian}
\end{equation}
where $F_{\mu \nu}=\partial_\mu A_\nu-\partial_\nu A_\mu$ is the field strength; $D_\mu=\partial_\mu+ig A^{\mathrm{ext}}_\mu+ig A_\mu$ is the covariant derivative and $m$ is the mass of the charged particle. We assume the scalar self-coupling, i.e. $\lambda(\phi\phi^*)^2/4$, is sufficiently small that we may ignore it, at least in the range of energies considered. Note that for QED no such term would arise.

We write the false vacuum transition amplitude as a path integral and note that we may integrate out the charged particle, as it enters quadratically.
\begin{align}
 \langle\Omega|\hat{S}_E|\Omega\rangle &= \int \mathcal{D}A_\mu\mathcal{D}\phi\ \mathrm{e}^{-\int_x\mathcal{L}_{\mathrm{SQED}}} \nonumber \\
 &= \int \mathcal{D}A_\mu \det(-D^2+m^2)^{-1}\ \mathrm{e}^{- \int_x \frac{1}{4}F^{\mu \nu}F_{\mu \nu}} \nonumber \\
 &= \int \mathcal{D}A_\mu \ \mathrm{e}^{-\mathrm{Tr}\log(-D^2+m^2) - \int_x \frac{1}{4}F^{\mu \nu}F_{\mu \nu}}, \label{eq:pathintegral1}
\end{align}
where $\int_x:=\int \mathrm{d}^4x$ and the functional integrations are normalised such that the amplitude is 1 for zero external field. The normalisation drops out once we take the imaginary part of the logarithm to find the rate, as in Eq. \eqref{eq:rate_imag}.

We can now use Schwinger's trick (i.e. Frullani's integral) to express the logarithm as a proper time integral \cite{schwinger1951gauge}
\begin{equation}
 \log(A)=-\int_0^\infty\frac{\mathrm{d}s}{s}(\mathrm{e}^{-As}-\mathrm{e}^{-s}),
\end{equation}
and drop the second term as it is field independent and will not contribute an imaginary part. The UV divergences of the theory will then turn up as divergences at small $s$ which can be renormalised using the heat kernel expansion. Introducing the proper time integral leads to the expression $\mathrm{Tr}(\mathrm{e}^{-(-D^2+m^2)s})$, which we express as a path integral over closed worldlines, \cite{feynman1951operator,gelfand1959integration,polyakov1980gauge,polyakov1988two,fradkin1991path}
\begin{equation}
 \mathrm{Tr}(\mathrm{e}^{-(-D^2+m^2)s}) = \int\mathcal{D}x^\mu \mathrm{e}^{-S_0[x^\mu,A^{\mathrm{ext}}_\mu+A_\mu;s]},
\end{equation}
where the action is given by
\begin{equation}
 S_0[x^\mu,a_\mu;s]:= m^2 s+ \frac{1}{4}\int^s_0\mathrm{d}\tau\dot{x}^\mu\dot{x}_\mu-ig\oint a_\mu \mathrm{d}x^\mu, \label{eq:s0_definition}
\end{equation}
and $\dot{x}^\mu:=\mathrm{d}x^\mu/\mathrm{d}\tau$. This is the worldline path integral for a charged scalar particle with the reparameterisation invariance fixed such that the einbein (also called the vierbein or tetrad by analogy to 4 dimensions) is equal to 2 (see for example chapter 1 of \cite{polchinski1998string}). The false vacuum transition amplitude is now
\begin{align}
 \Gamma &= -\frac{2}{\mathcal{V}}\mathrm{Im}\log\int \mathcal{D}A_\mu \mathrm{e}^{- \int_x \frac{1}{4}F^{\mu \nu}F_{\mu \nu}} \Bigg\{1+\nonumber \\
 &\sum_{n=1}^\infty\frac{1}{n!}\bigg(\prod\limits_{j=1}^{n} \int_0^\infty\frac{\mathrm{d}s_j}{s_j} \int\mathcal{D}x_j^\mu \mathrm{e}^{-S_0[x_j^\mu,A^{\mathrm{ext}}_\mu+A_\mu;s_j]}\bigg)\Bigg\}.
\end{align}
 At each order in $n$ the integration over the photon is now Gaussian and can be done exactly, resulting in an effective non-local worldline action. If we denote the free photon propagator by $G_{\mu\nu}(x_j,x_k)$, we can write this as
  \begin{widetext}
\begin{equation}
 \Gamma = -\frac{2}{\mathcal{V}}\mathrm{Im}\log \Bigg\{ 1+\sum_{n=1}^\infty\frac{1}{n!}\bigg(\prod\limits_{j=1}^{n} \int_0^\infty\frac{\mathrm{d}s_j}{s_j} \int\mathcal{D}x_j^\mu \bigg)\mathrm{e}^{-\sum_{k=1}^n (S_0[x_k^\mu,A^{\mathrm{ext}}_\mu;s_k] -\frac{g^2}{2}\sum_{l=1}^n \oint\oint \mathrm{d}x_k^\mu \mathrm{d}x_l^\nu G_{\mu\nu}(x_k,x_l))} \Bigg\} . \label{eq:amp_no_approx} 
\end{equation}
\end{widetext}
Integrating out the photon has left us with a non-local, long range interaction. At this point we have made no approximations regarding the strength of the external field, or of the coupling. The relatively simple exponential form of Eq. \eqref{eq:amp_no_approx} only obtains for Abelian gauge fields (see for example \cite{dotsenko1979renormalizability}).

At weak coupling, $g^2\ll 1$, the non-local interaction term in Eq. \eqref{eq:amp_no_approx} can be dropped at leading order. In this case the sum exponentiates, leaving only one path integration which can be carried out exactly, leading to Schwinger's result \cite{schwinger1951gauge}
\begin{equation}
\Gamma_{\mathrm{Schwinger}} = \frac{m^4\epsilon^2}{8\pi^3} \sum^\infty_{n=0}\frac{(-1)^{n+1}}{n^2} \mathrm{e}^{-\frac{\pi }{\epsilon}n}. \label{eq:schwinger}
\end{equation}
In this paper we consider arbitrary coupling, $g$, for which the non-local interaction cannot be dropped.

\subsection{Finite temperature rate\label{sec:finite_temp_rate}} 

The derivation thus far has been at zero temperature. At finite temperature, $T=1/\beta$, we make the following replacements
\begin{align}
\langle\Omega|\hat{S}_E|\Omega\rangle &\to \mathcal{N}^{-1}\mathrm{Tr} \ \mathrm{e}^{-\hat{H}\beta}, \nonumber \\
\mathcal{V} &\to \mathcal{V}_T, \nonumber
\end{align}
where $\hat{H}$ is the Hamiltonian of the system in the presence of the external field and the normalisation, $\mathcal{N}^{-1}$, ensures the amplitude is 1 in the absence of the external field (see \cite{gavrilov2008one} for a physical discussion of thermal Schwinger pair creation). In the second line $\mathcal{V}_T$ is equal to the spatial volume, $V$, multiplied by the inverse temperature $\beta$\footnote{Eq. \eqref{eq:rate_thermal} for the thermal rate has been advocated by Linde \cite{linde1981fate,linde1983decay}. An analysis by Langer shows that a different expression for $\mathcal{V}_T$ should be used, with the inverse temperature replaced by the decay time of an intermediate state \cite{langer1969statistical,affleck1981quantum,arnold1987sphalerons}. Though, as we only work to exponential accuracy (i.e. the leading order of the logarithm) in this paper, the difference does not affect our results.}. The rate is then given by
\begin{equation}
\Gamma_T = \frac{2}{\mathcal{V}_T} \mathrm{Im}\bigg\{ -\log\left(\mathrm{Tr}\  \mathrm{e}^{-\hat{H}\beta}\right)\bigg\}.  \label{eq:rate_thermal}
\end{equation}
This transition to finite temperature can be made straightforwardly using the Matsubara formalism, i.e. by enforcing periodicity in the Euclidean time coordinate, $x^4(\tau)= x^4(\tau)+\beta$, and including interactions between the periodic copies.

Including interactions between periodic copies is equivalent to replacing the photon propagator, $G_{\mu\nu}(x_j,x_k)$, by its thermal cousin, $G_{\mu\nu}(x_j,x_k;T)$. In a general $R_\xi$ gauge the $\xi$ dependent term drops out when integrated around a closed loop leaving just a term proportional to $\delta_{\mu\nu}$. This gauge independent part is
\begin{align}
G_{\mu\nu}(x_j,x_k;T)&:= \sum_{n=-\infty}^{\infty}G(x_j,x_k+\frac{n}{T} e_4) \delta_{\mu\nu}\nonumber \\
&=\sum_{n=-\infty}^{\infty}\frac{-\delta_{\mu\nu}}{4\pi^2(x_j-x_k-\frac{n}{T} e_4)^2} \nonumber \\
&=\frac{T \sinh (2 \pi T r_{jk})\ \delta_{\mu\nu}}{4 \pi  r_{jk} \left(\cos \left(2 \pi  T t_{jk}\right)-\cosh (2 \pi T r_{jk})\right)}, \label{eq:thermalsum}
\end{align}
where $e_4$ is the unit vector in the Euclidean time direction and we have defined $t_{jk}:=x_j^4-x_k^4$ and $r_{jk}:=\sqrt{(x_j^1-x_k^1)^2+(x_j^2-x_k^2)^2+(x_j^3-x_k^3)^2}$. This is the Matsubara thermal Green's function in position space.

To generalise Eq. \eqref{eq:amp_no_approx}, and get an expression for the rate at finite temperature, one need only replace the zero temperature Green's function with that of Eq. \eqref{eq:thermalsum}, and impose periodic boundary conditions in the Euclidean time direction, with period $1/T$. The aim of this paper is to calculate this thermal rate.

\subsection{Inclusive rate at fixed energy}

We will also consider inclusive tunnelling rates at a fixed energy $\mathcal{E}$, i.e. rates from a microcanonical ensemble. In this case one makes the replacements
\begin{align}
 \langle\Omega|\hat{S}_E|\Omega\rangle &\to \mathcal{N}^{-1}\mathrm{Tr}\left( \delta(\mathcal{E}-\hat{H}) \right), \nonumber \\
 \mathcal{V} &\to \mathcal{V}_{\mathcal{E}}, \nonumber
\end{align}
where the normalisation again ensures the amplitude is 1 in the absence of the external field. In the second line $\mathcal{V}_{\mathcal{E}}$ is equal to the spatial volume, $V$, multiplied by some timescale, which we expect to be $O(1/\mathcal{E})$ on dimensional grounds. The exact form of $\mathcal{V}_{\mathcal{E}}$ will not concern us in this paper as our final results are only to exponential accuracy. The rate, $\Gamma_\mathcal{E}$, is then given by
\begin{equation}
\Gamma_{\mathcal{E}} = \frac{2}{\mathcal{V}_{\mathcal{E}}} \mathrm{Im}\bigg\{ -\log\left(\mathrm{Tr}\  \delta(\mathcal{E}-\hat{H})\right)\bigg\},  \label{eq:rate_inclusive}
\end{equation}
The thermal density matrix is related to the microcanonical one by a sum over Boltzmann weights, or a Laplace transform, 
\begin{equation}
\mathrm{e}^{-\hat{H}\beta} = \int_0^\infty \mathrm{d} \mathcal{E} \mathrm{e}^{-\mathcal{E}\beta}\delta({\mathcal{E}}-\hat{H}). \label{eq:thermal_microcanonical}
\end{equation}
Hence the inverse relation is via an inverse Laplace transform,
\begin{equation}
\delta({\mathcal{E}}-\hat{H})=\lim_{B\to\infty}\int_{-iB}^{iB} \frac{\mathrm{d}\beta}{2\pi i}\mathrm{e}^{(\mathcal{E}-\hat{H})\beta} .
\end{equation}
The microcanonical density operator is the projection operator onto the subspace of states with energy, $\mathcal{E}$. These rates and their relationship to thermal tunnelling rates have been discussed by various authors \cite{affleck1981quantum,selivanov1986tunneling,khlebnikov1991periodic,rubakov1992towards,tinyakov1992multiparticle}. For sufficiently slow rates one can expand the logarithms in Eqs. \eqref{eq:rate_thermal} and \eqref{eq:rate_inclusive} to derive the following approximate relation,
\begin{equation}
\Gamma_{T} \sim \int_0^\infty \mathrm{d} \mathcal{E}\mathrm{e}^{-\mathcal{E}\beta} \Gamma_{\mathcal{E}}, \label{eq:inclusive_rate}
\end{equation}
where we have ignored the ratio $\mathcal{V}_{\mathcal{E}}/\mathcal{V}_{T}$ as we will only use the relation to exponential accuracy.

\section{The dilute instanton gas} \label{sec:approximation}

We wish to consider Schwinger pair production in QED and SQED for arbitrary coupling, $g$. This requires going beyond perturbation theory in $g$. For a sufficiently weak external field, as we will show, an alternative set of approximations are valid and allow us to proceed. These are the semiclassical and dilute instanton gas approximations.

\begin{figure}
 \centering
  \includegraphics[width=0.48\textwidth]{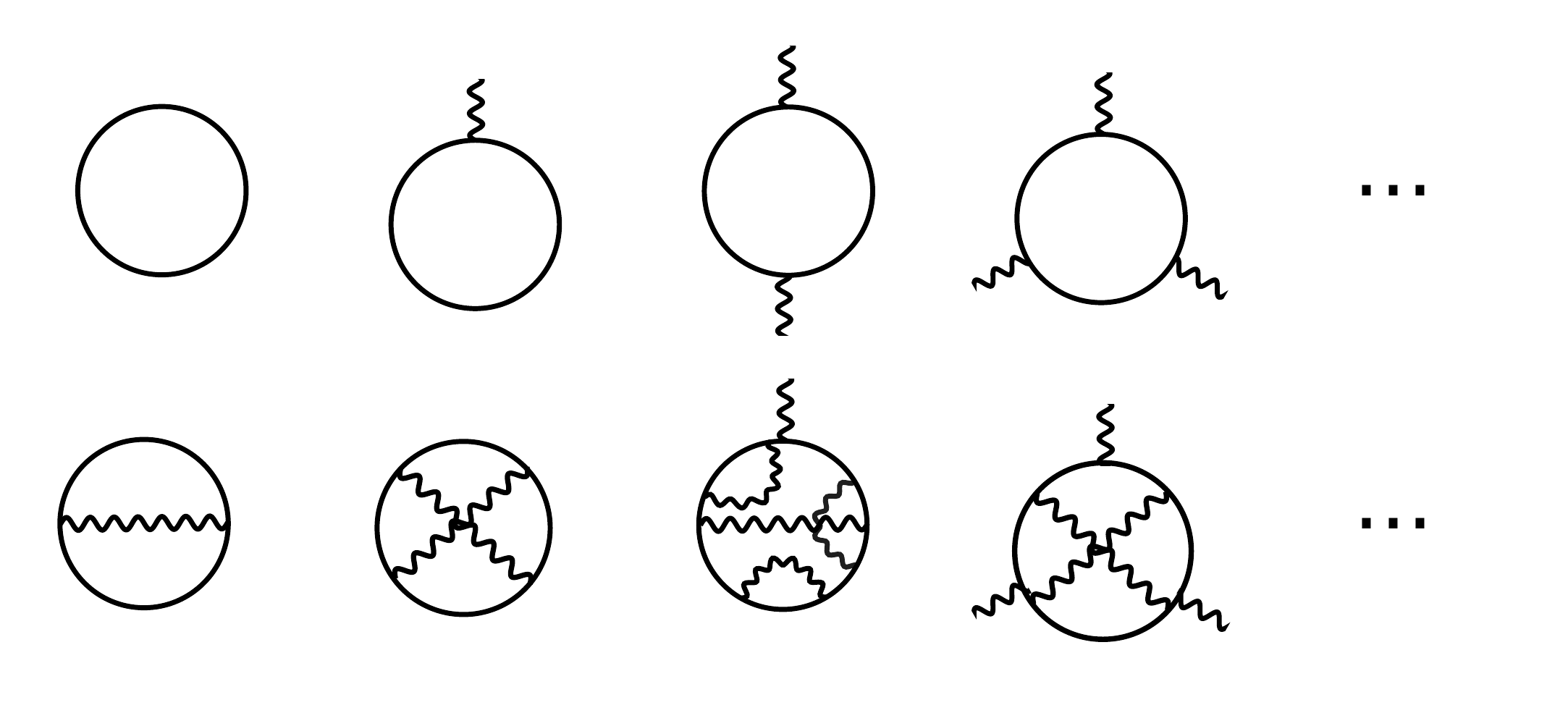}
  \caption{External photon legs denote couplings to the fixed external field whilst internal photon lines denote dynamical virtual photons. The Schwinger formula (Eq. \eqref{eq:schwinger}), valid at weak coupling, accounts for the infinite set of diagrams represented in the first row. The quenched approximation also includes all diagrams which include any number of internal photon lines, with any topology. Some examples are shown in the second line. Note that in SQED there are also four-point interactions involving two photons and two charged particles (not shown here though included in quenched approximation). In all the diagrams there is only one charged particle loop.}
  \label{fig:quenched3}
\end{figure}

Although Feynman diagrams will not be utilised in this calculation, they can illuminate the structure of the approximations we will make. The rate, Eq. \eqref{eq:amp_no_approx}, contains only connected Feynman diagrams, due to the logarithm. The constituents of the contributing diagrams are internal charged particles lines; external photon lines, for $A^{\mathrm{ext}}_\mu$; internal, dynamical photon lines, for $A_\mu$, and vertices joining two charged particle lines and one or two photon lines.

At weak coupling, $g\ll 1$, to leading order all dependence on the dynamical photon can be dropped. The path integrations in Eq. \eqref{eq:amp_no_approx} are then uncoupled and the sum exponentiates. The Feynman diagrams which contribute to this all contain one charged particle loop and an arbitrary number of external photon lines. These are the diagrams in the first row of Fig. \ref{fig:quenched3}. The sum of these diagrams at zero temperature is Schwinger's original result, Eq. \eqref{eq:schwinger}. At finite temperature the rate has been calculated in Refs. \cite{cox1984finite,elmfors1994electromagnetic,elmfors1998thermally,gies1999qed}. The inclusion of a single dynamical photon line (i.e. two loops) was calculated first by Ritus at zero temperature \cite{ritus1975lagrange,ritus1977connection,lebedev1984virial,ritus1998effective} and by Gies at finite but low temperature \cite{gies2000qed}. In these calculations the approximation of weak external fields has not been made.

At stronger coupling one must include the extra infinitely many diagrams containing arbitrary numbers of internal, dynamical photon lines as well as arbitrary numbers of charged particle loops. However, as we will argue, for a sufficiently weak external field, diagrams with a large number of charged particle loops will be suppressed and hence a loop expansion in charged particle loops is possible. At each order one must sum the infinite set of diagrams containing a fixed, finite number of charged particle loops and an arbitrary number of both external and dynamical photon lines. To first order this is the quenched approximation, which in this context was argued to be valid (at zero temperature) in the Refs. \cite{affleck1981monopole} and \cite{affleck1981pair} (see also \cite{gies2005quantum}). Fig. \ref{fig:quenched3} shows some examples of diagrams which contribute in the quenched approximation.

Following Refs. \cite{affleck1981monopole} and \cite{affleck1981pair}, we consider the situation where the external field is weak and sufficiently slowly varying to be considered constant. We choose the external field to point in the 3 direction, $F_{\mu\nu}=-i(\delta_{\mu 3}\delta_{\nu 4}-\delta_{\nu 3}\delta_{\mu 4})E$ (the factor of $-i$ is present due to the Wick rotation and the fact that $E$ is the value of the Minkowskian field). As long as the worldline $x^\mu(\tau)$ forms the boundary of some surface within the space, we can use Stokes' theorem to re-express the interaction with the external field,
\begin{align}
 -ig\oint A^{ext}_\mu \mathrm{d}x^\mu &= -\frac{ig}{2}\int \int F_{\mu\nu}\mathrm{d}x^\mu \wedge \mathrm{d}x^\nu \nonumber \\
 &= -gE\oint x_3 \mathrm{d}x^4,
\end{align}
which is simply the area enclosed by the worldline, projected onto the 3-4 plane and multiplied by $-gE$. Now, we are in a position to set up the weak field approximation to Eq. \eqref{eq:amp_no_approx}, which will amount to a semiclassical approximation. To see this it will be useful for us to rescale the $\tau$ in the integrand of $S_0$, the parameters $s_j$ and the fields $x^\mu_j(\tau)$. We rescale them according to
\begin{align}
 \tau &\to \tau/s_j, \nonumber \\
 s_j &\to s_j/gE, \nonumber \\
 x^\mu_j &\to x^\mu_jm/gE, \label{eq:scaling}
\end{align}
making all three dimensionless. The inverse temperature must be scaled in the same way as $x^\mu_j(\tau)$. We define the scaled temperature $\Temp:=mT/gE$ and $\tilde{\beta}:=1/\Temp$.

The full rate at finite temperature becomes, upon rescaling,
\begin{widetext}
\begin{equation}
\Gamma_T = -\frac{2}{\mathcal{V}_T}\mathrm{Im}\log \Bigg[1+\sum_{n=1}^\infty\frac{1}{n!}\prod\limits_{j=1}^{n}\left( \int_0^\infty \frac{\mathrm{d}s_j}{s_j}\int\mathcal{D}x_j^\mu \ \mathrm{e}^{ -\frac{1}{\epsilon}  \Action[x_j;s_j;\kappa,\Temp]} \ \mathrm{e}^{ \frac{\kappa}{\epsilon} \sum_{k<j} \oint\oint \mathrm{d}x_j^\mu \mathrm{d}x_k^\nu G_{\mu\nu}(x_j,x_k;\Temp) }\right)\Bigg], \label{eq:rate_no_approx}
\end{equation}
\end{widetext}
where $\epsilon:=gE/m^2$ and $\kappa:=g^3 E/m^2$ and we have defined, $\tilde{S}$, the scaled action as
\begin{align}
 \Action&[x;s;\kappa,\Temp] := s + \frac{1}{4s}\int^1_0\mathrm{d}\tau\dot{x}^\mu\dot{x}_{ \mu}-\int_0^1\mathrm{d}\tau x_{ 3} \dot{x}^4 \nonumber \\
 &- \frac{\kappa}{2}\int_0^1\mathrm{d}\tau\int_0^1\mathrm{d}\tau' \dot{x}^\mu(\tau) \dot{x}^\nu(\tau') G_{\mu\nu}(x(\tau),x(\tau');\Temp). \label{eq:scaled_action_s}
\end{align}
Everything inside the logarithm is now separately dimensionless. Note that Eq. \eqref{eq:rate_no_approx} is exact. We have as yet made no approximations regarding the strength of the external field or the coupling.

The parameters $m$, $g$, $E$ and $T$ only arise in equations \eqref{eq:rate_no_approx} and \eqref{eq:scaled_action_s} in three combinations: in the overall prefactor of the exponent, $1/\epsilon$, as $\kappa$ and as $\Temp$. For a sufficiently weak external field, $\epsilon\ll 1$, the path integral is calculable in the stationary phase, or semiclassical, approximation. This is independent of the value of the coupling, $g$. The worldline configurations which dominate the path integral are those which satisfy the classical equations of motion. Of these, those which give a non-zero imaginary part are those which are saddle points of the action with an odd number of negative modes in the spectrum of fluctuations about the solution. The solutions relevant to tunnelling have just one negative eigenvalue and are called bounces or instantons.

Note that the requirement that $\epsilon\ll 1$, which ensures semiclassicality, entails that
\begin{align}
\kappa &\ll g^2 \nonumber, \\
\Temp &\gg \frac{T}{m} \label{eq:kappa_temp_requirements},
\end{align}
however, as we make no restrictions on $g$ or $T$, $\kappa$ and $\Temp$ are not thuswise constrained. This is key as we will only calculate to leading order in $\epsilon$ but to all orders in $\kappa$ and $\Temp$.

\begin{figure}
 \centering
  \includegraphics[width=0.48\textwidth]{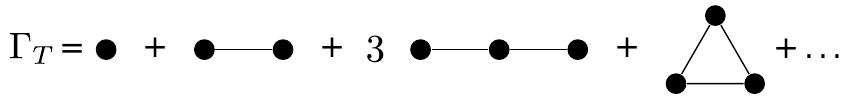}
  \caption{The first three orders of the cluster expansion of the rate. Each circle symbolises a closed worldline. The lines joining them are interactions given by the two particle function of Eq. \eqref{eq:fkl}. These diagrams are expressed algebraically in equations \eqref{eq:cluster_expansion} and \eqref{eq:cluster_first_3}.}
  \label{fig:clustering_expansion}
\end{figure}

To proceed in calculating the rate, Eq. \eqref{eq:rate_no_approx}, we perform a cluster expansion, as introduced by Ursell \cite{ursell1927evaluation} and Mayer \cite{mayer1937statistical}. We define the two-particle function $f_{kl}$, for $k\neq l$, by
\begin{equation}
 f_{kl} = \exp\bigg\{ \frac{\kappa}{\epsilon} \oint\oint \mathrm{d}x_k^\mu \mathrm{d}x_l^\nu G_{\mu\nu}(x_k,x_l;\Temp) \bigg\}-1. \label{eq:fkl}
\end{equation}
The cluster expansion to Eq. \eqref{eq:rate_no_approx} is then found by expanding in powers of $f_{kl}$ and grouping connected terms into so-called clusters. Only connected terms contribute to $\Gamma_T$. The expansion can be written as
\begin{equation}
 \Gamma_T = \sum_{n=1}^{\infty} \gamma_n, \label{eq:cluster_expansion}
\end{equation}
where $\gamma_n$ is the contribution to $\Gamma_T$ from clusters of $n$ worldlines. The terms in the expansion can be mapped to connected graphs of increasing complexity, such as in Fig. \ref{fig:clustering_expansion} (these are textbook results, see for example Ref. \cite{pathria2009statistical}). The $\gamma_n$ are proportional to the imaginary parts of what might conventionally be called cluster integrals (commonly denoted $b_n$) for the ensemble of charged particle worldlines, and so for brevity we will refer to them as cluster integrals.

The first three are given by
\begin{align}
 \gamma_1 &= -\frac{2}{1!\mathcal{V}_T}\mathrm{Im} \int_0^\infty \frac{\mathrm{d}s_1}{s_1}\int\mathcal{D}x_1^\mu \mathrm{e}^{-\frac{1}{\epsilon}\Action[x_1;s_1;\kappa,\Temp]}, \nonumber \\
  \gamma_2 &= -\frac{2}{2!\mathcal{V}_T}\mathrm{Im} \prod\limits_{j=1}^{2} \left(\int_0^\infty \frac{\mathrm{d}s_j}{s_j}\int\mathcal{D}x_j^\mu \mathrm{e}^{-\frac{1}{\epsilon}\Action[x_j;s_j;\kappa,\Temp]}\right) f_{12}, \nonumber \\
  \gamma_3 &= -\frac{2}{3!\mathcal{V}_T}\mathrm{Im} \prod\limits_{j=1}^{3} \left(\int_0^\infty \frac{\mathrm{d}s_j}{s_j}\int\mathcal{D}x_j^\mu \mathrm{e}^{-\frac{1}{\epsilon}\Action[x_j;s_j;\kappa,\Temp]}\right) \nonumber \\
  &\qquad\qquad\qquad\qquad\qquad\bigg\{3f_{12}f_{13}+f_{12}f_{13}f_{23}\bigg\}. \label{eq:cluster_first_3} 
\end{align}
Eq. \eqref{eq:cluster_expansion} is still formally exact but, importantly, is now expressed in a form that we can directly approximate. We follow Refs. \cite{polyakov1975compact,callan1977toward,callan1979theory,levine1978higher,jevicki1980statistical,ilgenfritz1979phase,bogomolny1980calculation} in performing a dilute instanton gas approximation. Essentially we will assume that the leading order behaviour of $\Gamma_T$ is captured by the lowest non-zero term in the cluster expansion. This is a self-consistent approximation, the higher order cluster integrals being exponentially suppressed with respect to the leading term.

First, suppose that there exists an instanton for $\gamma_1$, so that $\gamma_1\neq 0$. The path integral is invariant under translations $x^\mu(\tau)\to x^\mu(\tau)+a^\mu$. The instanton solution for $\gamma_1$ will necessarily break the translation symmetry and hence fluctuations around the instanton will contain (at least) four zero modes. Integration over these can be done using the collective coordinate method (see for example  \cite{polyakov1976quark}), resulting in an integral over the position of the instanton.

The higher order cluster integrals give the contributions due to interactions between instantons. Approximate multi-instanton solutions can be constructed by combinations of single instantons a large distance apart. The contribution of these approximate saddle points can be found using the method of constrained instantons \cite{levine1978higher,affleck1981constrained}. The integrations over the collective coordinates and constraints of these approximate instantons will take the form of cluster integrals for a gas of classical point particles (rather than worldlines), with dipole interactions (as the worldlines are closed and hence have zero net charge). In this way, the infinite number of degrees of freedom of each particle worldline are reduced to the four degrees of freedom of a point in spacetime.

From this perspective, the rate, $\Gamma_T$, can be interpreted as the pressure of the instanton gas. Standard statistical mechanical relations then give the density of instantons, $n_{\mathrm{inst}}$, as
\begin{equation}
n_{\mathrm{inst}} =  \sum_{n=1}^{\infty} n \gamma_n. \label{eq:density}
\end{equation}
Combining equations \eqref{eq:cluster_expansion} and \eqref{eq:density}, the rate, $\Gamma_T$, can be written as an expansion in powers of the density
\begin{equation}
\Gamma_T = n_{\mathrm{inst}} + B_2 n_{\mathrm{inst}}^2 + B_3 n_{\mathrm{inst}}^3 + O(n_{\mathrm{inst}}^4). \label{eq:virial}
\end{equation}
This is the Virial expansion and the coefficients, $B_n$, are the Virial coefficients. For $n\ge 3$ they are given by the irreducible graphs in the cluster expansion, those which cannot be cut into two pieces by cutting one line, and $B_2n_{\mathrm{inst}}^2 = -\gamma_2$. At weak coupling, this virial expansion has been introduced previously in Refs. \cite{lebedev1984virial,ritus1998effective}.

To leading order in the cluster expansion the instanton density will be given simply by $\gamma_1$. The average separation between instantons is then $\gamma_{1}^{-1/4}$. The density of instantons can be considered small if this distance is much larger than the maximum size of the instantons, $R$.

There is however a subtlety due to the long range interactions of the instantons which was also found in the dilute instanton gas expansion of QCD \cite{callan1977toward,callan1979theory}. The contribution to the action due to the interaction between a pair of dipoles in four dimensions a distance $|x|$ apart, decreases as $1/|x|^4$. This is such that, at zero temperature, its integral over the volume of spacetime diverges proportionally to $\log(\mathcal{V}_T)$. As a result there is such a divergence in the second Virial coefficient, $B_2$, and in all \emph{reducible} diagrams, defined to be those diagrams that can be split into two disconnected parts by cutting a single line. On the other hand, at non-zero temperature, there is no logarithmic divergence due to the finite extent of the Euclidean time direction.

Hence, at finite temperature, for sufficiently small, non-zero $\gamma_1$, we expect
\begin{equation}
 \Gamma_T = \gamma_1\left(1 + O(\gamma_1 L^4)\right), \label{eq:density_expansion_1}
\end{equation}
where $L=\mathrm{Max}(R,\beta)$.  This leading order approximation is equivalent to the quenched approximation. The semiclassical approximation of $\gamma_1$ gives $\gamma_1 L^4\sim \mathrm{e}^{-\Action(\kappa,\Temp)/\epsilon}$, where we have written $\Action(\kappa,\Temp)$ for the value of the scaled action, $\Action[x;s;\kappa,\Temp]$, evaluated at the saddle point.

On the other hand, if there does not exist an instanton solution consisting of a single worldline, then $\gamma_1=0$. In this case we must repeat the above arguments for the first non-zero cluster integral, $\gamma_{n_0}$, say. In that case the particles of the instanton gas would consist of groups of $n_0$ worldlines and Eq. \eqref{eq:density_expansion_1} would be replaced by
\begin{equation}
 \Gamma_T = \gamma_{n_0}\left(1 + O(\gamma_{n_0} L^4)\right).
\end{equation}

In this paper, we consider only the leading order term in the dilute instanton gas approximation, $\gamma_{n_0}$. Further, we only calculate the exponential suppression of the leading term. This is equivalent to saying that we calculate the logarithm of the rate to leading order in the small parameter $\epsilon$. When $n_0=1$, this is
\begin{equation}
 \log(\Gamma_T) = -\frac{\Action(\kappa,\Temp)}{\epsilon} + O(\log(\epsilon)).
\end{equation}
In semiclassically evaluating the terms $\gamma_n$, the saddle point of the $s_j$ integrations can be easily found. For $\gamma_1$, we find
\begin{align}
 \gamma_1 = -\frac{2 m^4\epsilon^4}{\tilde{\mathcal{V}_{\Temp}}}&\sqrt{2\pi \epsilon}\nonumber \\ 
 &\mathrm{Im}\int\mathcal{D}x^\mu  \bigg(\int^1_0\mathrm{d}\tau\dot{x}^\mu\dot{x}_\mu\bigg)^{-\frac{1}{4}}
 \mathrm{e}^{-\frac{1}{\epsilon}\Action[x;\kappa,\Temp]} \label{eq:amp_weak_field} 
\end{align}
where $\tilde{\mathcal{V}}_{\Temp}:= \mathcal{V}_Tm^4 \epsilon^4$ is the (dimensionless) scaled volume and we have defined $\Action[x;\kappa,\Temp]$ to be the scaled action evaluated at the saddle point of the $s$ integration,
\begin{equation}
 \Action[x;\kappa,\Temp] := L[x]-A[x]+\kappa V[x;\Temp],\label{eq:geometric_action}
\end{equation}
written in this way to emphasise its geometric nature. The constituent terms are the (parameterisation fixed) length of the worldline
\begin{equation}
L[x]:=\sqrt{\int^1_0\mathrm{d}\tau\dot{x}^\mu\dot{x}_{\mu}}, \label{eq:gaugefixedlength}
\end{equation}
the area projected onto the 3-4 plane
\begin{equation}
A[x]:=\int \mathrm{d}\tau x_{3} \dot{x}^4, \label{eq:area}
\end{equation}
and the interaction term
\begin{align}
&V[x;\Temp]:=\nonumber \\
&\frac{1}{2} \int_0^1\int_0^1 \mathrm{d}\tau\mathrm{d}\tau ' \dot{x}^\mu(\tau) \dot{x}^\nu(\tau ') G_{\mu\nu}(x(\tau),x(\tau ');\Temp).
\end{align}
The first term, $L[x]$, is the only non-geometric term, in the sense that it depends on the coordinates along the worldline. It is however equal to the length of the worldline when evaluated on-shell. Note that the action is invariant under $\tau\to\tau + c$, where $c$ is a constant. The corresponding conserved charge is $\dot{x}^2(\tau)$.

In some cases there may be no instanton solution consisting of a single worldline. As we have argued, in these cases one should next look for instanton solutions consisting of two and then more worldlines. The (scaled) action for $n_0$ worldlines could be thought of as that for a single discontinuous worldline (where one does not take derivatives across the discontinuities), except that the kinetic term, Eq. \eqref{eq:gaugefixedlength}, does not appear to be additive. However, the kinetic term is, in fact, additive if each of the disconnected worldlines have the same (parameterisation fixed) length,
\begin{align}
 &n_0\sqrt{\int^1_0\mathrm{d}\tau\dot{x}_1^\mu\dot{x}_{1\mu}} = \nonumber \\
 &\sqrt{\int^{1/n_0}_0\mathrm{d}\tau\dot{x_1}^\mu\dot{x}_{1\mu}+\dots +\int^1_{(n_0-1)/n_0}\mathrm{d}\tau\dot{x}_{n_0}^\mu\dot{x}_{n_0\mu}}.
\end{align}
For the instanton solutions relevant in this paper, this allows us to always talk about a single (possibly discontinuous) worldline and to always use the action in Eq. \eqref{eq:geometric_action}.

\section{Instantons\label{sec:instantons}} 

\subsection{Finite temperature rate}

The problem of finding the rate of pair production due to a weak external field at given $g$, $E$, $m$ and $T$ is now reduced to a problem which depends only on two parameters, $\kappa$ and $\Temp$. The general solution amounts to finding the saddle point of $\Action[x;\kappa,\Temp]$ with one negative mode, and the fluctuations about it.

Integrations over fluctuations in the negative mode, via an analytic continuation, give the all important factor of $i$  \cite{langer1967theory,langer1969statistical,coleman1977fate,callan1977fate}. There are also zero modes due to translation invariance. Integration over these degrees of freedom requires first introducing a constraint which fixes the translation invariance and then integrating over that constraint. We choose to fix the centre of mass of the worldline to be at the origin, $\bar{x}^\mu=0$. Integration over the constraint then gives a factor $\tilde{\mathcal{V}}_T=m^4\epsilon^4\mathcal{V}_{\Temp}$, cancelling the $1/\tilde{\mathcal{V}}_{\Temp}$ in Eq. \eqref{eq:amp_weak_field}. The remaining integrations over positive mode fluctuations give a subleading prefactor.

To calculate the logarithm of the rate to leading order in $\epsilon$, we need only find the instanton solution and calculate its action, $\Action(\kappa,\Temp)$. Even this is a difficult enough problem, made so by the non-local photon interaction in \eqref{eq:geometric_action}. In the following we consider the equations of motion analytically in certain limits: $\kappa \ll 1$, $\Temp\ll 1$ and large $\Temp$. Then for arbitrary $\kappa$ and $\Temp$ we use numerical methods.

\subsection{Inclusive rate at fixed energy}

From $\Action(\kappa,\Temp)$, we can calculate the inclusive rate at fixed energy. In the semiclassical approximation, Eq. \eqref{eq:inclusive_rate} shows that the two rates are related by a Laplace transform and hence the exponents of the rates are related via a Legendre transform. In the thermodynamic language $\Action(\kappa,\Temp)$ is the free energy divided by the temperature. The (scaled) energy of the solution $\Erg$ is
\begin{equation}
 \Erg=\frac{\partial \Action}{\partial \tilde{\beta}}, \label{eq:erg_def}
\end{equation}
corresponding to a physical energy $\mathcal{E}=m\Erg$. By further scaling the worldlines by $x\to x/\beta$, taking the derivative with respect to $\tilde{\beta}$, and then reversing the scaling, we find the following useful result
\begin{equation}
 \tilde{\beta} \Erg = L[x]-2A[x], \label{eq:erg}
\end{equation}
which holds on-shell. The exponential suppression of the rate of pair production at fixed energy is
\begin{align}
 \Sigma&=\frac{1}{\epsilon}(\Action-\Erg\tilde{\beta}) \nonumber \\
 &=\frac{1}{\epsilon}\Entropy(\kappa,\Erg).
\end{align}

\subsection{Regularisation\label{sec:regularisation}} 

As we have mentioned the interaction term, $V$, diverges at zero distance. For smooth worldlines, this is the long known self-energy divergence of electromagnetism. Its appearance in the worldline formulation of QED has been studied by many authors (see for example \cite{polyakov1980gauge,dotsenko1979renormalizability,brandt1981renormalization,karanikas1992infrared}). The divergence, being due to the strong interactions between nearby sections of a worldline, is proportional to its length.

We first consider a well known regularisation scheme due to Polyakov \cite{polyakov1980gauge}. At zero temperature this amounts to replacing the interaction term, $V[x;0]$, with
\begin{align}
 V_{\mathrm{Polyakov}}[x;0] &:= \frac{1}{8\pi^2} \int_0^1\int_0^1\mathrm{d}\tau\mathrm{d}\tau' \frac{\dot{x}^\mu(\tau) \dot{x}_\mu(\tau')}{(x(\tau)-x(\tau'))^2+a^2}\nonumber \\
 &-\frac{1}{8\pi^2} \frac{\pi}{a}\int_0^1 \sqrt{\dot{x}^2(\tau)}\mathrm{d}\tau. \label{eq:regularisation_polyakov}
 \end{align}
The second term in \eqref{eq:regularisation_polyakov}, proportional to the length of the worldline, is a counterterm which absorbs the short distance divergence of the first term. It is almost of the same form as the term $L[x]$ in the action (Eq. \eqref{eq:gaugefixedlength}), except without the reparameterisation fixing. On-shell the two terms are equal, hence we can see it as a mass counterterm.

This self-energy divergence has been shown to be the only divergence for smooth loops with no intersections \cite{dotsenko1979renormalizability}. Worldlines with discontinuous first derivatives (cusps) and intersections may arise when there are delta function interactions in the action or when a ratio of scales is taken to zero. Such worldlines also generate logarithmic divergences\footnote{These logarithmic divergences can be interpreted as due to Brehmstrahlung radiation. They give the anomalous dimension for Wilson loops, and hence for the propagator of the charged particles.}.

Unfortunately the regularisation scheme of Eq. \eqref{eq:regularisation_polyakov} leads to problems when trying to formulate the equations of motion which prevent us taking the limit $a \to 0$ in our numerical calculations. This is because, for sufficiently small $a$, the counterterm gives a negative bare mass. To bypass this problem we adopt an alternative regularisation in our numerical calculations for which the bare mass and the renormalised masses are equal,
 \begin{align}
  &V_R[x;0]  := \frac{1}{8\pi^2} \int_0^1\int_0^1\mathrm{d}\tau\mathrm{d}\tau' \frac{\dot{x}^\mu(\tau) \dot{x}_\mu(\tau')}{(x(\tau)-x(\tau'))^2+a^2}\nonumber \\
 &-\frac{1}{8\pi^2} \frac{\sqrt{\pi}}{a^2}\int_0^1\int_0^1\mathrm{d}\tau\mathrm{d}\tau' \dot{x}^\mu(\tau) \dot{x}_\mu(\tau')\mathrm{e}^{-(x(\tau)-x(\tau'))^2/a^2}.\label{eq:regularisation_rajantie}
 \end{align}
Equations \eqref{eq:regularisation_polyakov} and \eqref{eq:regularisation_rajantie} agree as $a \to 0$ as can be seen by recognising the Gaussian representation of the delta function. At finite temperature we must include the infinite sum of interactions with the periodic copies. If the periodic copies are disconnected and a finite distance apart, there is no ultraviolet divergence from their interaction and the unregularised interaction may be used. However, if the periodic copies are connected, the interaction between them must be regularised.

As well as the mass, there is, of course, charge renormalisation. Physically this is due to charged particle-antiparticle pairs popping into and out of existence and screening the bare charge. Thus the dilute instanton gas approximation, which only takes into account a small number of charged particle loops, does not take into account these effects. In the worldline formalism such short-lived virtual pairs are represented by small, closed worldlines. Though there are many such possible fluctuations, any given one will have an action of order $\epsilon^0$ and hence will not arise in the stationary phase approximation we have made. Including these fluctuations should result in the final rates depending on the renormalised charge as argued for in Ref. \cite{affleck1981pair}.

For the small worldlines of the short-lived virtual pairs to simply renormalise the charge, there must be a separation of scales between them and the larger worldlines which constitute the saddle point. We can make a simple estimate for the scale of the virtual pairs by equating the rest mass to the Coulomb attraction. This equality reads $2m=g^2/(4\pi r)$ and gives the distance between charges as $r=g^2/(8\pi m)$. In our dimensionless units, this translates to a distance $\kappa/(8\pi)$, which must be smaller than any scale present in the instanton for the charge renormalisation effects to be independent. The modification of the photon-charged particle interaction at distances below $g^2/(8\pi m)$ has been discussed in Refs. \cite{goebel1970spatial,goldhaber1982monopoles}, with regard to magnetic monopoles.

\subsection{Small \texorpdfstring{$\kappa$}{kappa} expansion\label{sec:finitecoupling}} 

\subsubsection{A singular perturbation problem\label{sec:singular_perturbation}} 

Throughout this paper, we make the approximation that the external field is weak, i.e. that $0<\epsilon\ll 1$. The parameter $\kappa:=g^2 \epsilon$ is proportional to $\epsilon$. Hence, for not too large couplings, we will also have that $0<\kappa \ll 1$. This is the case we will consider in this section. Parametrically large couplings, such that $\kappa=O(1)$ will be considered in sections \ref{sec:lowtemp}, \ref{sec:hightemp} and \ref{sec:numerical_results}.

For sufficiently small $\kappa$, one would naively expect that we could simply set $\kappa=0$ in the scaled action $\Action[x^\mu;\kappa,\Temp]$, so dropping the interaction term. The problem is that the interaction term diverges at short distance and hence cannot be ignored for arbitrarily small but positive $\kappa$. This signals that for small $\kappa$ we are dealing with a singular perturbation problem related to the existence of widely separated scales (see for example \cite{bender1999advanced}).

\begin{figure}
 \centering
  \includegraphics[width=0.2\textwidth]{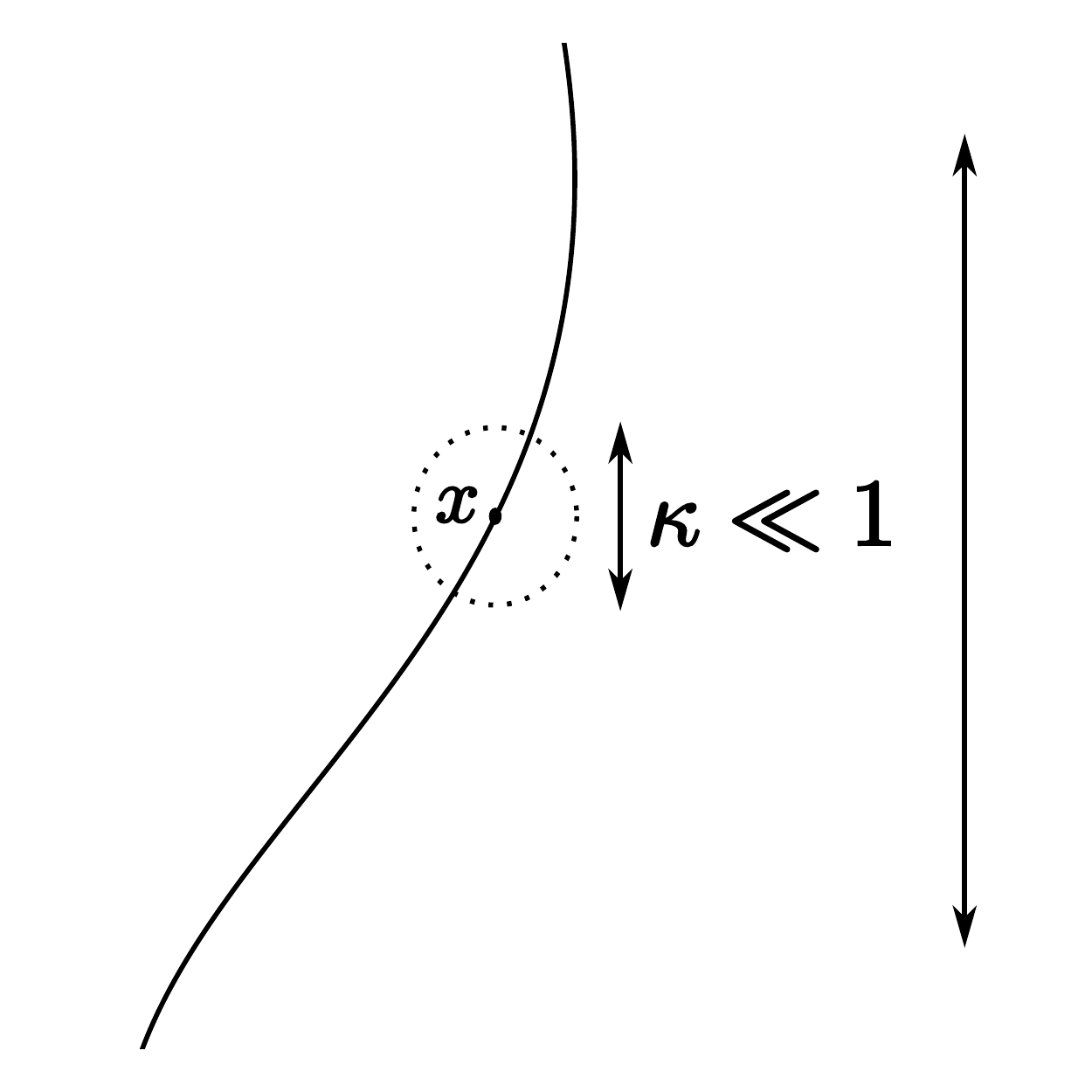}
  \caption{Sphere of influence of point $x$, size $O(\kappa$), compared to curvature of worldline on scale $O(1)$.}
  \label{fig:interaction_scale}
\end{figure}

We seek \emph{distinguished limits} of the action $\Action[x;\kappa,\Temp]$, and its corresponding equations of motion, by considering the scalings $x = \kappa^\alpha y$ and $\Temp = \kappa^{-\alpha} \Theta$. The aim is to find scalings such that there is a balance between two terms in the action and that the leading order equations of motion give non-trivial solutions. After the scaling the action is
\begin{equation}
\Action[x;\kappa,\Temp]=\kappa^\alpha L[y] - \kappa^{2\alpha} A[y] + \kappa V[y,y';\Theta].
\end{equation}
There are three distinguished limits: $\alpha=0,\ 1/2,\ 1$. The first, the $\alpha=0$ scaling, corresponds to scales $x= O(1)$, which we will refer to as the infrared (IR) problem. The last, the $\alpha=1$ scaling, corresponds to shorter scales $x=O(\kappa)$, which we will refer to as the ultraviolet (UV) problem. The intermediate scaling, $\alpha=1/2$ corresponds to scales $x= O(\kappa^{1/2})$, which we will refer to as the matching problem.

For small $\kappa$, an approximate solution to the equations of motion valid on all scales can be found by solving the leading order equations of motion in these three distinguished limits and matching them smoothly together. The simplest of the three problems is the matching problem, $\alpha=1/2$. The leading approximation amounts to simply keeping the length term
\begin{equation}
\Action(\kappa \ll 1,\Temp) \approx \kappa^{1/2} L[y]. \label{eq:alphahalf}
\end{equation}
The area and interaction terms are equally subdominant on these scales, both being of order $\kappa$. Solutions to the minimisation of the length term are simply straight lines. Hence the IR and UV solutions must be matched with straight lines. The matching is done at some scale $\lambda=O(\kappa^{1/2})$ which acts as a UV cut-off for the IR problem and as an IR cut-off for the UV problem. The final solution should be independent of the specific choice of $\lambda$.

The IR problem, the $\alpha=0$ scaling, in the leading approximation amounts to simply dropping the interaction term, i.e. to
\begin{equation}
\Action(\kappa \ll 1,\Temp) \approx L[y] - A[y]. \label{eq:alpha0}
\end{equation}
In terms of a Feynman diagram language, this approximation takes into account all external field photon exchanges but no virtual photon exchanges. This is the top row of Fig. \ref{fig:quenched3}, a one-loop approximation. Making this action stationary is the old problem of maximising the area of a field given a fixed length of fencing. The solution at zero temperature is a circle of radius 1 in the 3-4 plane. At finite temperature the solution can be found using the method of images.

The UV problem, the $\alpha=1$ scaling, in the leading approximation amounts to dropping the area term, i.e. to
\begin{equation}
\Action(\kappa \ll 1,\Temp) \approx \kappa (L[y] + V[y,y';\Theta]) \label{eq:alpha1}.
\end{equation}
This equation determines the dynamics at scales $y=\kappa^{-1} x=O(1)$. In terms of Feynman diagrams this approximation takes into account all virtual photon loops but no external photon lines.

\begin{figure}
 \centering
  \includegraphics[width=0.48\textwidth]{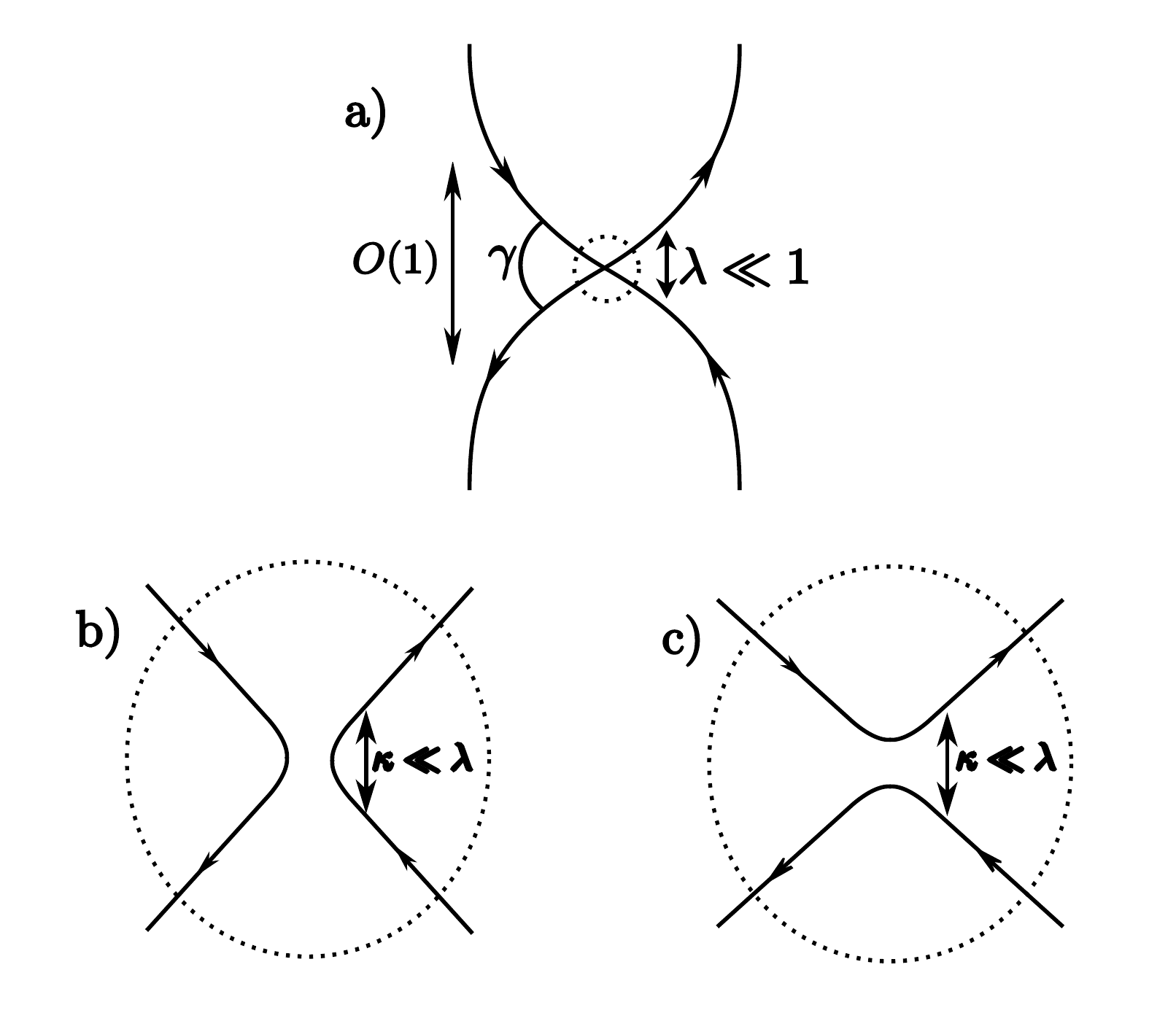}
  \caption{On scales $x=O(1)$ there may appear an intersection, as in a). This must be resolved on shorter scales, $x=O(\kappa)$, as b) or c). Due to the fact that the problem can be specified completely in the plane, these are the only possibilities. The long and short distance pictures are matched at some scale $\lambda=O(\kappa^{1/2})$.}
  \label{fig:intersection}
\end{figure}

Eq. \eqref{eq:alpha1} is the action of a massive charged particle, in the absence of an external field. Hence we can immediately find one solution, that of a straight line, the particle sitting still (or 4D rotations thereof). To find a solution to the full equations of motion, valid at all scales we can stitch this straight line solution together with a solution of Eq. \eqref{eq:alpha0}. This is possible if for every point $x$ on the worldline, we can draw a ball of size $\lambda$ within which the worldline appears straight as $\kappa \to 0$. As solutions to \eqref{eq:alpha0} are independent of $\kappa$, this will always be possible as long as the worldline has everywhere finite curvature and does not self-intersect. Otherwise, in the region of a cusp or self-intersection, the straight line solution to Eq. \eqref{eq:alpha1} cannot be used.

Such cusps or self-intersections are not permissible when all the parameters of the theory are finite as they give new divergences which depend on the angle at the non-analytic point, $\gamma$ in Fig. \ref{fig:intersection}. Due to this the divergences cannot be absorbed as counterterms in the field theory. However, if a ratio of scales in the problem goes to zero, as when $\kappa \to 0$, such apparent cusps or self-intersections can appear at larger scales, $O(1)$ and $O(\kappa^{1/2})$ in our case. On the smallest scale, $O(\kappa)$ in our case, these apparent non-analytic points can be resolved as in Fig. \ref{fig:interaction_scale}. The separation of scales in processes which involve a large momentum transfer such as deep inelastic scattering, gives divergences for the same reason \cite{korchemsky1985loop,gellas1996greens}. 

For apparent non-analytic points to be resolved on scales $O(\kappa)$, there must exist solutions of \eqref{eq:alpha1} which have the topology (in the plane) of b) and c) in Fig. \ref{fig:interaction_scale}. Note that these two possibilities are related by a rotation by $\pi/2$. Hence all points of self-intersection can be resolved if solutions with the topology of b) can be found for all angles $\pi/2<\gamma\leq \pi$. These solutions must be stitched together with the IR solutions at some scale $\lambda$. Thus we must impose boundary conditions at $\lambda$ such that the solutions can be smoothly matched. 

The existence of such solutions can be made plausible by noting that for $\pi/2<\gamma\leq \pi$ the scalar product between the tangent vectors on the left and right hand sides of b) is negative and hence the interaction term is repulsive. The magnitude of the repulsion increases without bound as the worldlines approach each other, suggesting that the worldlines should approach to some minimum distance, $|y-y'|=O(1)$. The minimum distance is a function of the incoming angle $\gamma$ and is independent of $\kappa$, as long as $\gamma$ and $\pi-\gamma$ are both $O(1)$. Naively one might expect that the scaled action $L[y]+V[y,y']$ would then be $O(1)$ and hence the contribution to the $\Action=\kappa( L[y]+V[y,y';\Theta])= O(\kappa)$. However the following argument shows that this is not the case.

The matching of the IR and UV solutions is carried out at the scale $\lambda$. Due to the long range of the interaction, the UV scaled action, $L[y]+V[y,y';\Theta]$, will get large as $\log(\lambda/\kappa)$. This is the infrared divergence of the interaction of two long straight worldlines which are not parallel. Upon matching the IR and UV solutions all dependence on $\lambda$ must drop out. However we will be left with a contribution to the action of the order $\kappa\log(\kappa)$.

Overall we find that for solutions with apparent cusps or intersections
\begin{equation}
 \Action(\kappa,\Temp) = c(\Temp) + d(\Temp) \kappa \log(\kappa) + O(\kappa),
\end{equation}
for some $c(\Temp)$ and $d(\Temp)$. For solutions without cusps or intersections the $\kappa\log(\kappa)$ term is absent, i.e. $d(\Temp)=0$.

\subsubsection{Small \texorpdfstring{$\kappa$}{kappa} results\label{sec:finite_coupling_results}} 

At zero temperature the solution to the IR problem is a circle of radius 1. At every point, $x$, on the circle, a small ball of radius $\lambda=O(\kappa^{1/2})$ can be drawn within which the worldline looks approximately straight. Hence, the circle of radius 1 solves the equations of motion at all scales. The resulting action is
\begin{equation}
\Action(\kappa,0) = \pi - \frac{\kappa}{4}. \label{eq:s_zerotemp}
\end{equation}
This result was first derived in \cite{affleck1981monopole,affleck1981pair}. Due to the symmetry of the problem this result is in fact exact for arbitrary $\kappa$ and hence applies even for parametrically strong coupling. The prefactor is given by the determinant of fluctuations about this solution. This can be computed at leading order in $\kappa$ giving for the rate
\begin{equation}
 \Gamma \approx (2s+1)\frac{m^4\epsilon^2}{8\pi^3}\ \mathrm{e}^{-\frac{1}{\epsilon}(\pi-\frac{\kappa}{4})},\label{eq:gamma_zerotemp}
\end{equation}
where $s$ is the spin of the charged particle. At $\kappa=0$ this reduces to Schwinger's result for weak external fields (Eq. \eqref{eq:schwinger}).

\begin{figure}
 \centering
  \includegraphics[width=0.48\textwidth]{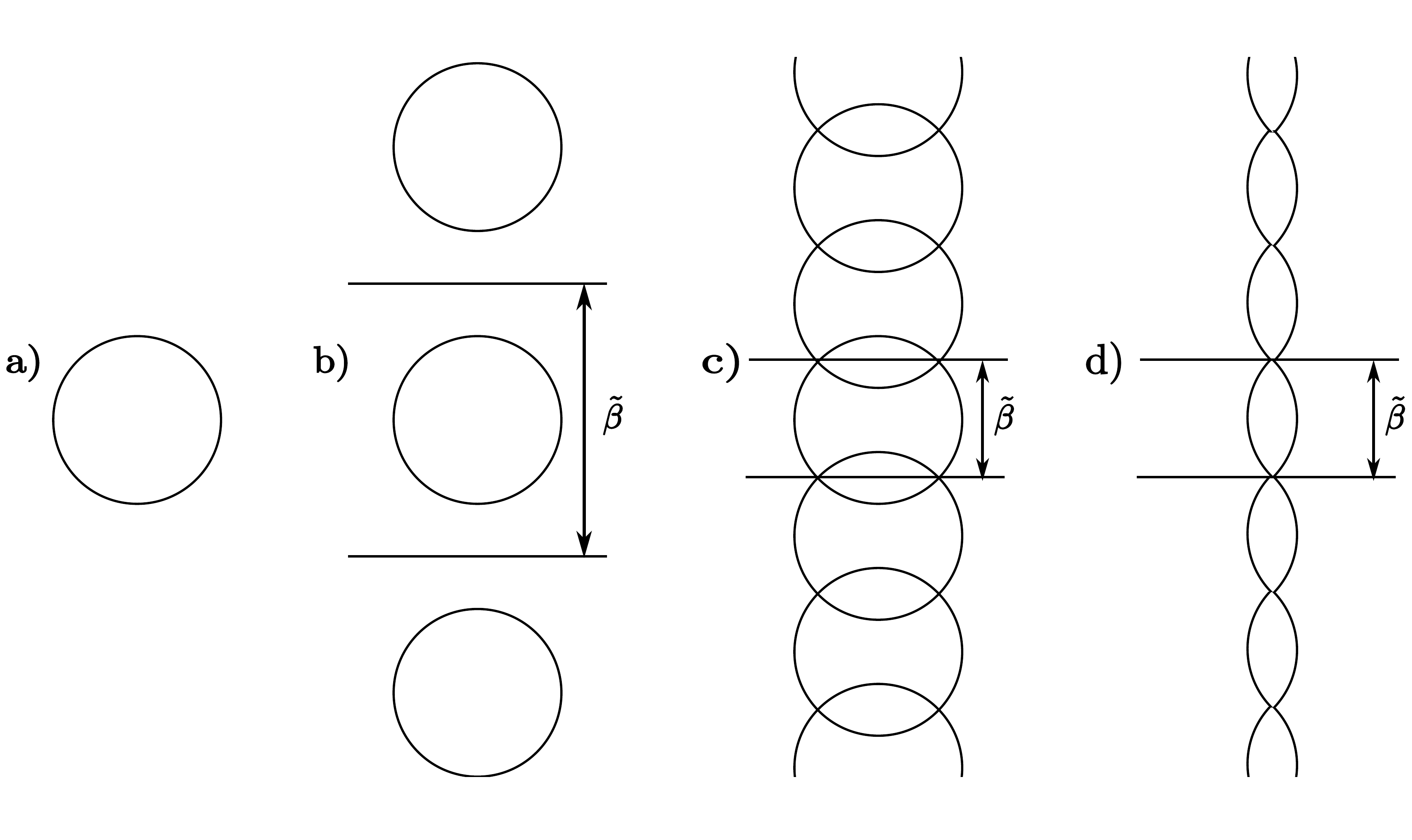}
  \caption{Instanton solutions at small $\kappa$: a) at zero temperature; b) at $0<\Temp<1/2$; c) at $\Temp>1/2$, the naive instanton of overlapping circles and d) at $\Temp>1/2$, the lemon instanton.}
  \label{fig:zeroCoupling}
\end{figure}

At finite temperatures, $\Temp$, and small $\kappa$, the leading order solution to the IR problem is given by an infinite sequence of circles of radius 1 separated by a distance $\tilde{\beta}$ along the Euclidean time axis (see Fig. \ref{fig:zeroCoupling} b)). For temperatures such that $\Temp < 1/2$, these circles do not overlap and, for sufficiently small $\kappa$, we are able to draw a small ball of radius $O(\kappa^{1/2})$ within which there is a single worldline which looks approximately straight. Hence for such temperatures, the sequence of circles solves the equations of motion at all scales and, to lowest order in $\kappa$, the rate is the same as at zero temperature. This means that, at one-loop order, we find no corrections to the zero temperature rate for $\Temp < 1/2$.

Corrections to this can be calculated using perturbation theory, for small $\kappa$. We write the full action and it's solution as expansions in $\kappa$,
\begin{align}
 \Action[x] &= \Action_0[x]+\kappa \Delta\Action[x], \nonumber \\
 x^\mu(\tau) &= x_0^\mu(\tau)+\kappa x_1^\mu(\tau)+\kappa^2 x_2^\mu(\tau)+\dots \label{eq:perturbation_theory_kappa}
\end{align}
where $\Action_0[x]=L[x]-A[x]$ and $\Delta\Action[x]=V[x]$. First order perturbation theory requires us to simply evaluate $\Delta\Action[x_0]$. We split this up into interactions between pairs of loops,
\begin{equation}
 \kappa\Delta\Action[x_0] = \kappa\sum_{n=-\infty}^{\infty} \Delta\Action_n[x_0], \label{eq:action_sum}
\end{equation}
where we have defined $\Delta\Action_n[x_0]$ to be the interaction between the loop at the origin and that centred at Euclidean time $n/\Temp$,
\begin{equation}
 \kappa \Delta\Action_n[x_0] = \frac{\kappa}{8\pi^2}\oint\oint\frac{\mathrm{d}x\mathrm{d}x'}{(x-x'-\frac{n}{\Temp} e_4)^2}.
\end{equation}
The $x$ denotes the positions on the circle at the origin with respect to the origin. The $x'$ denotes the positions on the circle centred at Euclidean time $n/\Temp$ with respect to its centre. The $e_4$ is a unit vector in the Euclidean time direction. The result of the integration, for $n\neq 0$, is
\begin{equation}
  \kappa \Delta\Action_n[x_0] = -\frac{\kappa}{4}\frac{\left(\sqrt{\frac{1}{4}-\left(\frac{\Temp}{n}\right)^2}-\frac{1}{2}\right)^2}{\sqrt{\frac{1}{4}-\left(\frac{\Temp}{n}\right)^2}}. \label{eq:lowtempcorrection}
\end{equation}
This was first derived in Ref. \cite{monin2010photon}. For $n = 0$, the integral is $-\kappa/4$ after regularisation and is the zero temperature correction in Eq. \eqref{eq:s_zerotemp}.
The full first order correction is given by the sum, \eqref{eq:action_sum}. It is negative for $\Temp\geq 0$ hence it increases the rate of pair production. It also diverges to $-\infty$ as $\Temp\to 1/2$, i.e. where the zero temperature instantons touch. However, following the discussion of section \ref{sec:singular_perturbation}, the separation of scales breaks down when neighbouring circles are only a distance $O(\kappa^{1/2})$ apart, when $1/2-\Temp = O( \kappa^{1/2})$.

Unlike the zero temperature result, there are corrections at second order in $\kappa$ due to the warping of the shape of the circles. To calculate these, we must solve
\begin{equation}
 \int_0^1 \left( \frac{\delta^2 \Action_0}{\delta x^\mu(\tau) \delta x^\nu(\tau')}\bigg|_{x_0}x_1^\nu(\tau') \right) \mathrm{d}\tau' + \frac{\delta \Delta\Action}{\delta x^\mu(\tau)}\bigg|_{x_0}=0. \label{eq:second_order_eom}
\end{equation}
The solution must lie in the $3-4$ plane, due to the symmetry of the problem, and must be closed. Hence, we may express it as
\begin{equation}
 x_1^\mu(\tau)=\epsilon(\tau)(0,0,\cos(2\pi\tau),\sin(2\pi\tau)).
\end{equation}
In terms of $\epsilon(\tau)$, the terms in the action at second order in $\kappa$ are
\begin{align}
&\frac{\kappa^2}{2}\left(\int_0^1  \frac{\dot{\epsilon}^2}{2\pi}\mathrm{d}\tau -2\pi\bar{\epsilon}^2\right)\nonumber \\
&- \frac{\kappa^2\Temp^4}{2\pi}\int_0^1\left(4\zeta(4)+ 24 \Temp^2 \zeta(6) (1-\cos (4 \pi  \tau ))\right)\epsilon(\tau)\mathrm{d}\tau, \label{eq:action_second_order}
\end{align}
where $\bar{\epsilon}^2$ denotes the square of the average of $\epsilon(\tau)$ (not the average of the square) and $\zeta$ denotes the Riemann zeta function\footnote{Note that had the kinetic term been the actual length, rather than its reparameterisation fixed form, the only difference would be the replacement of $\bar{\epsilon}^2$ with $\int_0^1\epsilon(\tau)^2\mathrm{d}\tau$.}. From Eq. \eqref{eq:action_second_order} we find the equations of motion for $\epsilon(\tau)$, the solution of which can be found straightforwardly via a Fourier series expansion. The two arbitrary parameters in the general solution are fixed by satisfying the constraint, $\bar{x}^\mu=0$, which we are using to fix translation invariance.
The solution is given by
\begin{equation}
 \kappa\epsilon(\tau) = -\frac{\kappa \Temp^4}{2\pi}\bigg\{ 4\zeta(4) + 24 \zeta(6)\Temp^2 + 6 \zeta(6)\Temp^2\cos(4\pi \tau)\bigg\}. 
\end{equation}
The constant terms reduce the radius of the circle and the term proportional to $\cos(4\pi\tau)$ makes the circle prolate (stretched in the $x^4$ direction). Substituting this solution into Eq. \eqref{eq:action_second_order} and putting it together with the zeroth and first order terms we arrive at
\begin{widetext}
\begin{equation}
  \Action(\kappa,\Temp) = \pi -\frac{\kappa}{4}\Bigg\{1 +2\sum_{n=1}^{\infty}  \frac{\left(\sqrt{\frac{1}{4}-\left(\frac{\Temp}{n}\right)^2}-\frac{1}{2}\right)^2}{\sqrt{\frac{1}{4}-\left(\frac{\Temp}{n}\right)^2}}\Bigg\} + \frac{\kappa^2}{\pi}(4 \zeta(4)^2 \Temp^8 +48 \zeta(4) \zeta(6) \Temp^{10} + 126 \zeta(6)^2 \Temp^{12} ) + O(\kappa^3). \label{eq:action_kappa_second_order}
\end{equation}
\end{widetext}
For the corresponding inclusive rate at fixed energy we consider the Legendre transform of this sum. The energy is, from Eq. \eqref{eq:erg_def},
\begin{equation}
\Erg=-\Temp^2\frac{\partial\Action(\kappa,\Temp)}{\partial\Temp}.
\end{equation}
The leading term on the right hand side takes the form of $\kappa$ multiplying a function of $\Temp$. A consideration of this function implies that if we wish to consider energies much larger than $\kappa$, the corresponding temperature must be very close to $1/2$. This is the region of parameter space where the circular worldlines almost touch, the minimum distance $d\ll 1$. The UV problem in this case is non-relativistic, with $y^3$ being the ``time'' direction. Solutions to this non-relativistic problem exist for $d\gtrsim \kappa^{1/2}$. This implies that the Legendre transform of Eq. \eqref{eq:action_kappa_second_order} is only valid for $\Erg\lesssim \kappa^{1/4}$. For $\Erg\lesssim \kappa$ taking the Legendre transform analytically is made difficult by the infinite sum. In the limited regime $\kappa \ll \Erg \ll \kappa^{1/4}$ however we can fairly simply find the leading few terms.
\begin{align}
&\Entropy(\kappa,\Erg) = \pi - 2 \Erg -\frac{3}{4}\kappa^{2/3} \Erg^{1/3}  \nonumber \\
&+ \frac{\kappa }{4}\bigg\{ 1 -\sum_{n=2}^{\infty}  \left(\sqrt{1-\frac{1}{n^2}}-1\right)^2 \left(1-\frac{1}{n^2}\right)^{-1/2} \bigg\} \nonumber \\ 
&-\frac{11}{64}\kappa^{4/3}\Erg^{-1/3} +\sum_{n=2}^{\infty}\frac{\kappa ^{5/3}\Erg ^{-2/3} }{32 n \left(n^2-1\right)^{3/2}} + \frac{35 }{1536 }\kappa^2\Erg ^{-1}\nonumber \\
&\qquad\qquad\qquad\qquad\qquad\qquad\qquad + O(\kappa^{7/3}\Erg^{-4/3}). \label{eq:entropy_small_kappa}
\end{align}
The leading enhancement, $-2\Erg$, has been long known in the context of induced vacuum decay \cite{affleck1979induced,voloshin1991illustrative}.

For $\Temp \geq 1/2$ the circles intersect and the above calculation breaks down (see c) in Fig. \ref{fig:zeroCoupling}). The intersection must be resolved in a region of size $\kappa$ as in Fig. \ref{fig:interaction_scale}. However, once we include intersections a more general class of solutions to the full problem are possible: we can combine sections of circles with intersections. This is possible as solutions to \eqref{eq:alpha0} must only locally be arcs of a circle with curvature 1. Of all possible solutions describing pair production processes, that with minimum action will dominate the path integral and hence give the rate of pair production.

The minimum action solution of this kind has been found by several authors \cite{selivanov1985destruction,selivanov1986tunneling,ivlev1987tunneling,brown2015schwinger}, though there is some dispute about this \cite{medina2015schwinger}. It is given by a lemon shape, Fig. \ref{fig:zeroCoupling} d), and not by the overlapping circles, c). The angle of intersection (see Fig. \ref{fig:intersection}) is given by
\begin{equation}
 \gamma_{\Temp}=2 \arcsin\left(\frac{1}{2\Temp}\right).
\end{equation}
However, note that this worldline on IR scales is only a solution to the full problem if, for given $\gamma_{\Temp}$, the corresponding UV solution exists. At small but finite $\kappa$ our numerical calculations in Sec. \ref{sec:numerical_results} find instanton solutions which appear to approach the lemon instanton as $\kappa\to 0$.

The action of the thermal lemon shaped instanton, to zeroth order in $\kappa$, is
\begin{equation}
\Action(0,\Temp) = \gamma_{\Temp} + \sin(\gamma_{\Temp}) \label{eq:s_lemon_kappa0}
\end{equation}
where $\Temp>1/2$. Below $\Temp=1/2$ the action is equal to $\pi$, at zeroth order in $\kappa$. The action and its first derivative are continuous at $\Temp=1/2$, though the second derivative diverges as $-8/\sqrt{T-1/2}$ as $\Temp\to 1/2$ from above. Hence we say that there is a second order phase transition at $\Temp=1/2$, for $\kappa=0$. Below the phase transition the solution has circular symmetry. This is broken above it.

To get the leading order correction in $\kappa$ it would seem we must solve Eq. \eqref{eq:alpha1} in the region of the intersection (assuming such a solution exists). However, we can in fact bypass this hard problem using perturbative renormalisation. First, we evaluate the interaction term on the leading order IR instanton, the lemon. This gives an unphysical logarithmic UV divergence from the intersections
\begin{equation}
V_{\mathrm{Lemon}} = \frac{\kappa}{\pi^2}\bigg\{\bigg(\frac{\pi}{2} - \gamma_{\Temp} \bigg) \cot (\gamma_{\Temp} )+1\bigg\}\log(\lambda) + \mathrm{finite \ terms}, \label{eq:vlemon}
\end{equation}
where we have used $\lambda$ for the short distance regulator, rather than $a$, as this should be of the order of the matching scale, as in section \ref{sec:singular_perturbation}. Though we cannot solve the UV problem, we know that it must provide a compensating counterterm i.e.
\begin{equation}
\frac{\kappa}{\pi^2}\bigg\{\bigg(\frac{\pi}{2} - \gamma_{\Temp} \bigg) \cot (\gamma_{\Temp} )+1\bigg\}\log\bigg(\frac{\kappa}{\lambda}\bigg).
\end{equation}
From the perspective of the short distance physics this is an IR divergence, arising due to the matching scale $\lambda$ being much larger than $\kappa$. The presence of $\kappa$ in the logarithm is due to the scaling in the short distance problem. This meets our expectations, as explained at the end of section \ref{sec:finitecoupling}, leading to a contribution of order $\kappa\log(\kappa)$,
\begin{align}
\Action(\kappa,\Temp) &= \gamma_{\Temp} + \sin(\gamma_{\Temp}) \nonumber \\
&+ \frac{1}{\pi^2}\bigg\{\bigg(\frac{\pi}{2} - \gamma_{\Temp} \bigg) \cot (\gamma_{\Temp} )+1\bigg\}\kappa \log(\kappa) + O(\kappa).\label{eq:themal_lemon} 
\end{align}
where $\Temp>1/2$. Note that the $O(\kappa\log(\kappa))$ term starts to dominate over the leading term when $\gamma_{\Temp} = O(\kappa)$ and $\pi-\gamma_{\Temp} = O(\kappa)$, or $\Temp=O(\kappa^{-1})$ and $\Temp-1/2=O(\kappa)$. This signals a breakdown of the separation of scales assumed in deriving Eq. \eqref{eq:themal_lemon} and a breakdown of the approximate solution.

We have perturbatively renormalised the problem. The subleading corrections at $O(\kappa)$ depend on the solution to the short distance ($\alpha=1$) problem. Our result could be nonperturbatively improved using the methods of the renormalisation group.

From Eq. \eqref{eq:themal_lemon} we can find the inclusive pair production rate at a fixed energy by Legendre transform,
\begin{align}
 \Entropy&(\kappa,\Erg)=\pi-\gamma_{\Erg} -\sin (\gamma_{\Erg} )\nonumber \\
 &+\frac{  1}{\pi^2}\bigg\{\bigg(\frac{\pi}{2} - \gamma_{\Erg} \bigg) \cot (\gamma_{\Erg} )+1\bigg\} \kappa\log (\kappa )+O(\kappa), \label{eq:entropy_zero_kappa}
\end{align}
where $\gamma_{\Erg}:=2 \arcsin(\Erg/2)$. The leading result here is the same as for pair production stimulated by the presence of a particle of mass $\Erg$ \cite{selivanov1985destruction} (or an off-shell photon \cite{torgrimsson2017dynamically}) or by a collision of particles with the same centre of mass energy \cite{rubakov1992initial,kiselev1992false}. These calculations involve the same shaped instanton, though without the periodic copies. The (scaled) exponential suppression in that case, including the $O(\kappa \log(\kappa))$ correction, is
\begin{align}
 \pi-&\gamma_{\Erg} -\sin (\gamma_{\Erg} )\nonumber \\
 &+\frac{1}{2\pi^2}\bigg\{ -\gamma_{\Erg} \cot(\gamma_{\Erg}) + 1\bigg\}\kappa \log(\kappa)+O(\kappa).
\end{align}
Note that as $0<\kappa<1$ and $|\gamma_{\Erg}|<\pi$, the corresponding rate is strictly lower than that given by the exponentiation of $\Entropy(\kappa,\Erg)$. This is as expected: the inclusive rate at energy $\Erg$ is greater that the rate of the specific process at the same energy.

\subsection{Low temperature expansion\label{sec:lowtemp}} 

Let us now consider the case of low temperatures, $\Temp\ll 1$, but arbitrary coupling, $\kappa$. The effect of non-zero temperature is felt through the interaction potential, coupling periodic copies of the circular worldline. Perturbation theory in $\Temp$ takes the form
 \begin{align}
 \Action[x] &= \Action_0[x]+\Temp^4 \Action_4[x] + \Temp^6 \Action_6[x] + \Temp^8 \Action_8[x] + \dots, \nonumber \\
 x^\mu(\tau) &= x_0^\mu(\tau)+\Temp^4 x_4^\mu(\tau)+\Temp^6 x_6^\mu(\tau)+\Temp^8 x_8^\mu(\tau)+\dots \label{eq:perturbation_theory_temp}
\end{align}
The terms $\Action_n[x]$ are simply defined to be the coefficients of $\Temp^n$ in the full action. Note that: i) there is no linear term in $\Action[x]$; ii) the coefficient of the quadratic term is zero due to the closure of the worldline loops and iii) the coefficients of all odd powers of $\Temp$ in $\Action[x]$ vanish due to cancellation between loops in the positive and negative Euclidean time directions. These properties of the $\Action[x]$ expansion carry over to that of $x^\mu(\tau)$ by standard perturbation theory.

First order perturbation theory gives
\begin{equation}
   \Action(\kappa,\Temp) = \pi-\frac{\kappa}{4}-\zeta(4)\kappa\Temp^4-4\zeta(6)   \kappa\Temp^6+O(\Temp^8). \label{eq:low_temp_leading2}
\end{equation}
These first two terms are the same as those coming from the expansion of Eq. \eqref{eq:action_kappa_second_order}. To calculate the coefficient of the $O(\Temp^8)$ term requires second order perturbation theory, which amounts to solving a somewhat complicated integrodifferential equation.

As pointed out in Refs. \cite{lebedev1984virial,monin2010photon}, expanding the exponential of Eq. \eqref{eq:low_temp_leading2} captures the two-loop corrections for weak fields and low temperatures (Eq. (76) in Ref. \cite{ritus1975lagrange} and Eq. (65) in Ref. \cite{gies2000qed}).
\begin{equation}
\mathrm{e}^{-\frac{1}{\epsilon}(\pi-\frac{\kappa}{4}-\zeta(4)\kappa\Temp^4)}\approx \left(1+\pi\alpha+\frac{2\pi^5\alpha T^4}{45m^4\epsilon^4}\right)\mathrm{e}^{-\pi/\epsilon},
\end{equation}
where $\alpha=g^2/(4\pi)$. At higher loop orders one would need to calculate also the semiclassical prefactor for comparison.

The Legendre transform of \eqref{eq:low_temp_leading2} gives the inclusive pair production rate at fixed, perturbatively low energies. It is
\begin{align}
 \Entropy(\kappa,\Erg)&=\pi -\frac{\kappa }{4}-\frac{5   \zeta (4)^{1/5} \kappa}{2^{8/5}}\left(\frac{\Erg}{\kappa}\right)^{4/5}\nonumber \\
 &-\frac{ \zeta (6) \kappa}{2^{2/5} \zeta (4)^{6/5}}\left(\frac{\Erg}{\kappa}\right)^{6/5} +O\left(\frac{\Erg}{\kappa}\right)^{8/5}
\end{align}
where the requirement of low temperatures translates into the requirement that $(\Erg/\kappa) \ll 1$.\footnote{The structure of this expansion is reminiscent of diagrammatic low energy expansions about instantons, such as arose in the discussion of electroweak baryon number conservation (see for example \cite{arnold1990baryon,voloshin1991illustrative}).}

\subsection{High temperature\label{sec:hightemp}} 

At sufficiently high temperatures the process of pair production becomes classical. The instanton is then independent of the Euclidean time direction and is called a sphaleron \cite{linde1981fate,linde1983decay,manton1983topology,klinkhamer1984saddle}. The problem reduces from four to three dimensions. Further, due to the irrelevance of the one and two directions ($\mu=1,2$) the problem becomes one dimensional. The action gives the Boltzmann factor. 

In our case the instanton consists of two worldlines: a stationary charged particle, $x^\mu(\tau)$, and its antiparticle, $y^\mu(\tau')$, at a fixed distance, $|x^3(\tau)-y^3(\tau')|=r$. On such a path the action reduces to
\begin{equation}
 \Action_{\mathrm{Straight}}[r;\kappa,\Temp] =  \bigg( 2 - r - \frac{\kappa}{4\pi r} \bigg) \frac{1}{\Temp}.
\end{equation}
There is a stationary point of the action at $r_0 = \sqrt{\kappa/4\pi}$ which gives the thermal instanton. The action is then
\begin{equation}
\Action_{\mathrm{Straight}}(\kappa,\Temp)=2 \bigg( 1 - \sqrt{\frac{\kappa}{4\pi}}\bigg) \frac{1}{\Temp}, \label{eq:action_straight}
\end{equation}
which can also be written as $\Action= \Erg/\Temp$, where $\Erg$ is the energy of the solution. Note that as $\Temp=\epsilon T/m$, the factors of $\epsilon$ cancel in the exponent of the rate leaving just the usual Boltzmann suppression in physical units.

The instanton may give the rate of thermal pair production when it is the lowest action solution for given parameters $(\kappa,\Temp)$. In a broad class of theories, though not including SQED or QED, it has been shown that a solution must satisfy a further constraint for it to describe tunnelling: the spectrum of linear perturbations about the solution must have one negative mode \cite{coleman1987quantum}. Variation of $r$ is one negative mode, present for all $\kappa$ and $\Temp$.

Due to the periodic boundary conditions, the components of both worldlines are each expressible as a Fourier series. We define $\zeta^\mu(\tau):=x^\mu(\tau)-y^\mu(\tau)$ and $\xi^\mu(\tau):=x^\mu(\tau)+y^\mu(\tau)$. The linearised eigenvalue equations about the straight line solution inherit the non-locality of the full action. They are thus linear integrodifferential equations, the expressions being very long, so we omit them here.

For $\kappa\ll 1$, the integrands of the non-local interactions become highly peaked, approaching delta functions which make the eigenvalue equations local. In this regime the dynamics is non-relativistic and the eigenvalue equations can be straightforwardly solved. In the spectrum of eigenfunctions there are two sets of potentially unstable linear fluctuations, one given by $\zeta^3(\tau)=r_0+\delta \cos(2\pi n\tau)$ and  the other by $\xi^3(\tau)=\delta \sin(2\pi n\tau)$, where $\delta\ll 1$, $n\in \mathbb{N}$, and in each case all other components are zero. The eigenvalues are
\begin{equation}
\lambda_n(\kappa,\Temp)\approx \frac{1}{2}(2 \pi n)^2 \Temp- \frac{4 \pi}{\sqrt{\pi\kappa }\Temp}. \label{eq:cos_eigenvalue_nonrel}
\end{equation}
The lowest frequency mode is thus least stable, and is unstable when $\Temp< \Temp_{\lambda_1=0}(\kappa)\approx\sqrt{2}\pi^{-3/4}\kappa^{-1/4}$. This signals the existence of another solution of lower action which is continuously connected to the straight line solution but which breaks time translation invariance. Hence there is a second order phase transition in the rate at this temperature. The instability is exactly analogous to the Plateau-Rayleigh instability in fluid dynamics \cite{plateau1873experimental,strutt1878instability}; the Gregory-Laflamme instability in black strings \cite{gregory1993black}; nuclear scission \cite{brosa1990nuclear} and an instability in vacuum bubbles at finite temperature \cite{linde1981fate,linde1983decay}. 

These linear fluctuations remain eigenvectors of the full integrodifferential equations at larger values of $\kappa$. The eigenvalue of the lowest frequency mode is then somewhat more complicated,
\begin{align}
\lambda_1(\kappa,\Temp)= \frac{1}{2}(2 \pi )^2 &\Temp-\frac{2}{3} \pi ^2 \kappa  \Temp^2- \frac{2 \pi}{\sqrt{\pi\kappa }\Temp} \nonumber \\
 &- 2 \pi\left( 1 + \frac{1 }{\sqrt{\pi \kappa}\Temp}\right)\mathrm{e}^{-\sqrt{\pi \kappa}  \Temp}. \label{eq:cos_eigenvalue}
\end{align}
For the higher frequency modes, the eigenvalue is given by $\lambda_n(\kappa,\Temp)=n\lambda_1(\kappa,n\Temp)$. The term, $-\tfrac{2}{3} \pi ^2 \kappa  \Temp^2$, is due to the interactions of each worldline with itself. It is the much-discussed, electromagnetic self-force \cite{dirac1938classical,wheeler1945interaction,gralla2009rigorous} and its contribution destabilises the straight worldlines, increasingly so at higher temperatures. This issue has been discussed in a similar context in  Ref. \cite{goebel1970spatial}, where it was argued that the photon-charged particle interaction is modified on sufficiently short scales, so diminishing the self-force.

\begin{figure}
 \centering
  \includegraphics[width=0.48\textwidth]{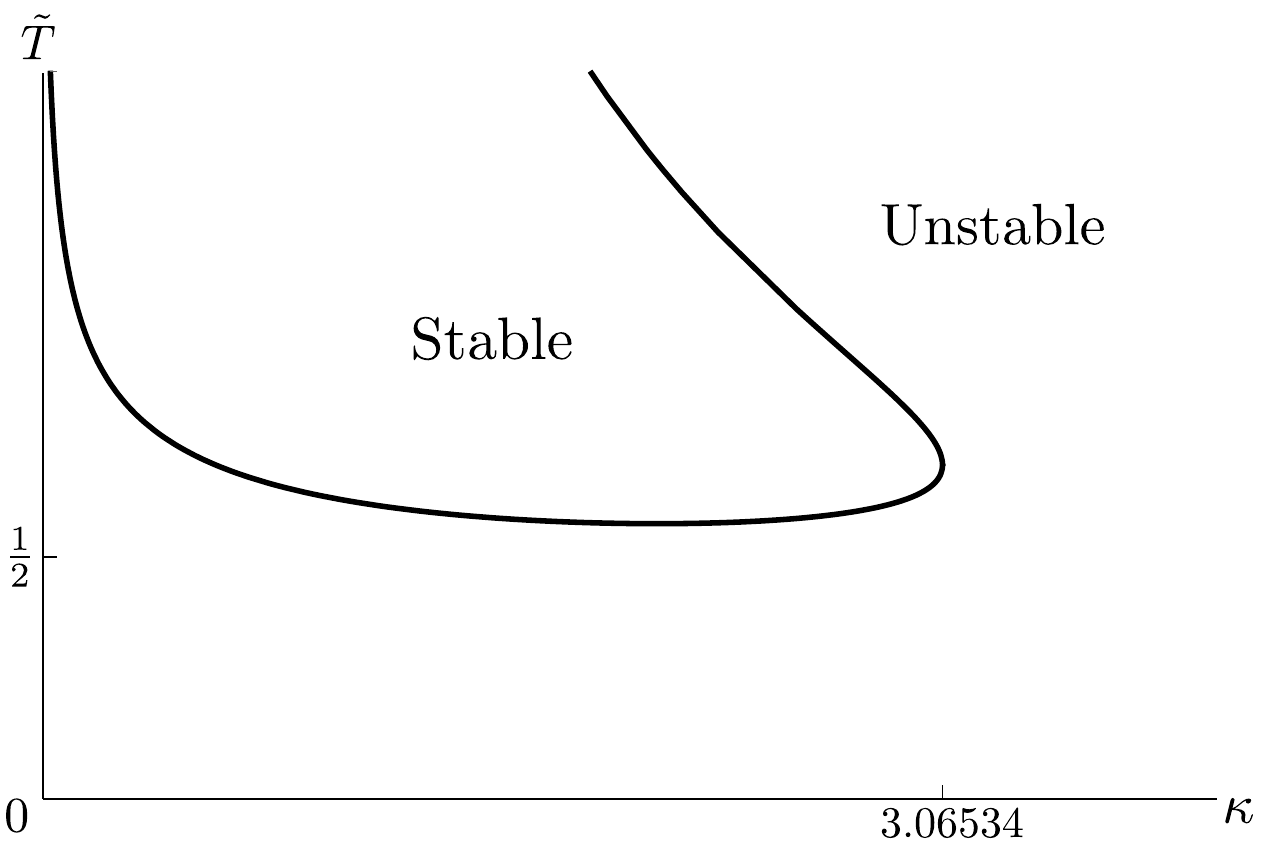}
  \caption{Regions within which the straight line instanton is stable and unstable to the lowest frequency perturbations. The boundary is defined by setting Eq. \eqref{eq:cos_eigenvalue} to zero and defines a function, $\Temp_{\lambda_1=0}(\kappa)$, with two branches.}
  \label{fig:instability}
\end{figure}

Setting Eq. \eqref{eq:cos_eigenvalue} to zero defines the boundary between stability and instability to the lowest frequency perturbation. The boundary is described by a function, $\Temp_{\lambda_1=0}(\kappa)$, consisting of two branches which meet at $\kappa\approx 3.06534$. This is shown in Fig. \ref{fig:instability}. For sufficiently small $\kappa$, the lower branch coincides with the long wavelength instability found in the non-relativistic analysis and the upper branch is approximately given by $\Temp_{\lambda_1=0}(\kappa)\approx 3/\kappa$. The instability above the upper branch is due to the self-force.

Higher frequency modes are more stable to the long wavelength instability but less stable to the self-force instability. As $n$ increases the self-force term grows fastest so all sufficiently high harmonics are unstable. This instability is present for all $\kappa$ and $\Temp$. In fact, as this instability only depends on the shape of the worldlines at short distances, $O(1/(n\Temp))$, it is present for all smooth worldlines (and likely for more general worldlines too). Due to the translational symmetry in the Euclidean time direction, the unstable harmonics of the sphaleron come in pairs, one a sine and the other a cosine.

The self-force instability may be a sign of the breaking down of the cluster expansion at larger values of $\kappa$. In support of this view, the self-force does not arise for $\kappa\ll 1$ at leading order in $\kappa$. In the cluster expansion, only a small number of charged particle worldlines are included. Charged particle loops of size $\kappa/(8\pi)$ have a small action, due to cancellation between the kinetic and interaction terms. These loops make up the bubbling sea of virtual charged particle pairs, part of the quantum vacuum. Their presence modifies the photon-charge interaction on scales of order $\kappa/(8\pi)$, an effect which is not included in the cluster expansion. A quantitative inclusion of these effects is beyond the scope of this paper but, as argued in Ref. \cite{goebel1970spatial}, photon-charge interactions should be weaker on scales of order $\kappa$ ($g^2/m$ in dimensionful units) and below. This is interpreted as an effective spreading-out of the charge, which prevents the self-
force instability.

For small deviations below the lower branch of $\Temp_{\lambda_1=0}(\kappa)$, and for sufficiently small $\kappa$, the dynamics is non-relativistic. In this case the equations of motion can be solved, even beyond the linearised approximation, by straightforward integration. The first integral of the motion is the energy, $\Erg$,
\begin{equation}
-\frac{1}{4}\dot{r}^2 + U(r) = \Erg,
\end{equation}
where overdot signifies differentiation with respect to the Euclidean time coordinate, $t$, and $U(r)=2 - r - \kappa/(4\pi r)$. Integrating this equation gives $t(r)$, which can be inverted to give $r(t)$. The solutions are periodic, with period $\tilde{\beta}(\kappa,\Erg)$.

In the non-relativistic regime the action is given by
\begin{equation}
\Action_{\mathrm{Nonrel}}(\kappa,\Temp) = \int_{r_L}^{r_R}\frac{2U(r)-\Erg}{\sqrt{U(r) - \Erg }}\mathrm{d}r ,
\end{equation}
where $\Erg$ is treated as a function of $\tilde{\beta}$ and $\kappa$ and where $r_L$ and $r_R$ are the classical turning points on the left and right. The non-relativistic approximation is valid for $\Delta \Temp:=\Temp_{\lambda_1=0}-\Temp\ll 1$, hence we expand the integral thuswise
\begin{align}
 \Action_{\mathrm{Nonrel}}(\kappa,\Temp)&=\Action_{\mathrm{Straight}}(\kappa,\Temp)-\frac{1}{3}  \sqrt{2} \pi ^{7/4} \kappa ^{5/4}\Delta\Temp^2\nonumber \\
 &-\frac{41}{54}  \pi ^{5/2} \kappa^{3/2}\Delta \Temp^3+O\left(\Delta \Temp^4\right),
\end{align}
Note that the non-relativistic solution has a lower action than the straight lines, so it dominates the rate where it exists,
\[
 \Action(\kappa,\Temp) =
  \begin{cases} 
      \hfill \Action_{\mathrm{Straight}}(\kappa,\Temp)  \hfill & , \Delta \Temp\leq 0, \\
      \hfill \Action_{\mathrm{Nonrel}}(\kappa,\Temp) \hfill & , 0<\Delta \Temp\ll 1. \\
  \end{cases}
\]
Also note that the difference between the two rates arises at second order in $\Delta \Temp$, showing that the transition between the two solutions is a second order phase transition.

The Legendre transform of these results give
\[
 \Entropy(\kappa,\Erg) =
  \begin{cases} 
      \hfill \Entropy_{\mathrm{Straight}}(\kappa,\Erg)  \hfill & , \Delta \Erg\leq 0, \\
      \hfill \Entropy_{\mathrm{Nonrel}}(\kappa,\Erg) \hfill & , 0<\Delta \Erg\ll 1. \\
  \end{cases}
\]
where $\Delta\Erg:=\Erg_c-\Erg$ and $\Erg_c:=2(1-\sqrt{\kappa/4\pi})$, the threshold energy. The two functions are
\begin{equation}
\Entropy_{\mathrm{Straight}}(\kappa,\Erg) = 0,
\end{equation}
and
\begin{align}
\Entropy_{\mathrm{Nonrel}}(\kappa,\Erg) &= \frac{\pi ^{3/4}  \kappa^{1/4}}{\sqrt{2}}\Delta \Erg+\frac{3 \pi ^{5/4} }{16 \sqrt{2} \kappa^{1/4}}\Delta \Erg^2\nonumber \\
&-\frac{5 \pi ^{7/4} }{256 \sqrt{2} \kappa ^{3/4}}\Delta \Erg^3+O\left(\Delta \Erg^4\right). \label{eq:sigma_nonrel}
\end{align}
The inclusive rate of pair production at a fixed energy is unsuppressed at the threshold energy. Just below the threshold, $\Delta\Erg\ll 1$, the suppression is given by the non-relativistic result here. Note that the leading term in $\Delta\Erg$ can be written as $\Delta\Erg/\Temp_{\lambda_1=0}$.

\section{Arbitrary temperature and \texorpdfstring{$\kappa$}{kappa}\label{sec:numerical_results}} 

For arbitrary temperature and $\kappa$ there is no symmetry and no small parameter which can help us proceed analytically.\footnote{Note that solutions of the equations of motion are only instantons if their actions are positive. At zero temperature this restricts us to $\kappa<4\pi$. The same condition holds at high temperatures for the straight line instanton.}

We adopt a numerical approach, in particular we discretise the loop, representing it by a large number, $N$, of points, $x_i$, $i=0,\dots,N-1$, and then write an approximation to the action where derivatives are replaced by finite differences (see Appendix \ref{appendix:numerics} for details). Note that this is not a lattice regularisation as the points are not constrained to lie on a lattice but may lie anywhere in $\mathbb{R}^4$, up to numerical accuracy.

The number $N$ must be chosen such that the distance between neighbouring points, $|dx_i|:=|x_{i+1}-x_i|$, is much smaller than the smallest scale in the problem, the cut-off, $a$. Note that for a continuous worldline, the global reparameterisation symmetry $\tau\to\tau+c$ means that $\dot{x}^2$ is constant. Thus, to leading order in $1/N$, $|dx_i|$ is independent of $i$ and hence equal to $L[x]/N$, where $L[x]$ is the length of the loop. Further, the cut-off $a$ must be chosen to be much smaller than any other scale in the problem. In summary we require
\begin{equation}
 \frac{L[x]}{N}\ll a \ll \mathrm{Min}(\kappa,A^{-1}[x;i]), \label{eq:scale_hierarchy}
\end{equation}
where $A[x;i]$ is the proper acceleration of the worldline at the point $i$. Note that the interactions between two disconnected worldlines do not need regularisation, so we may treat them exactly (up to discretisation errors). Computational constraints impose a maximum possible $N$ ($\sim 2^{12}$ in our case). This in turn imposes a minimum $a$ and hence a minimum $\kappa$ and a maximum proper acceleration.

The equations of motion are then $4 N$ (ignoring for the moment the symmetries and the zero modes) coupled, nonlinear algebraic equations which we solve numerically. Starting with an initial guess at the solution we iteratively solve the linearised equations until converging on a solution of the nonlinear equations, i.e. the Newton-Raphson method. An accuracy of better than $10^{-7}$ was usually reached in about 3 iterations. Simpler gradient methods cannot be used here as the solution is a saddle point, having one negative mode.

\subsection{Finite temperature results}

At low temperatures we can use the zero temperature instanton as an initial guess. Once the iterations have converged, we can then increase the temperature slightly and repeat the procedure using the solution from the last run as the initial guess for the next. In this way we can find all solutions in the $(\kappa,\Temp)$ plane which are continuously connected to the low temperature solutions. These all have the topology of a circle and we refer to them as C instantons.

There are also instantons with the topology of railway tracks: two infinitely long disconnected pieces. Over the whole $(\kappa,\Temp)$ plane there exist such solutions consisting of two straight lines (see Sec. \ref{sec:hightemp}) which we refer to as S instantons. Below the lower branch of $\Temp_{\lambda_1=0}$ (see Fig. \ref{fig:instability}), there exists another class of solutions with this topology and with lower action. These consist of two wavy lines and we refer to them as W instantons. For small $\kappa$ and for temperatures just below the lower branch of $\Temp_{\lambda_1=0}$, we can use the non-relativistic approximation of Sec. \ref{sec:hightemp} as an initial guess. From there we can step in $\kappa$ and $\Temp$ to find all the continuously connected solutions. As we approach the lower branch of $\Temp_{\lambda_1=0}$ from below, the W instantons become straighter and merge with the S instantons.

\begin{figure}
 \centering
  \includegraphics[width=0.48\textwidth,height=0.18\textwidth]{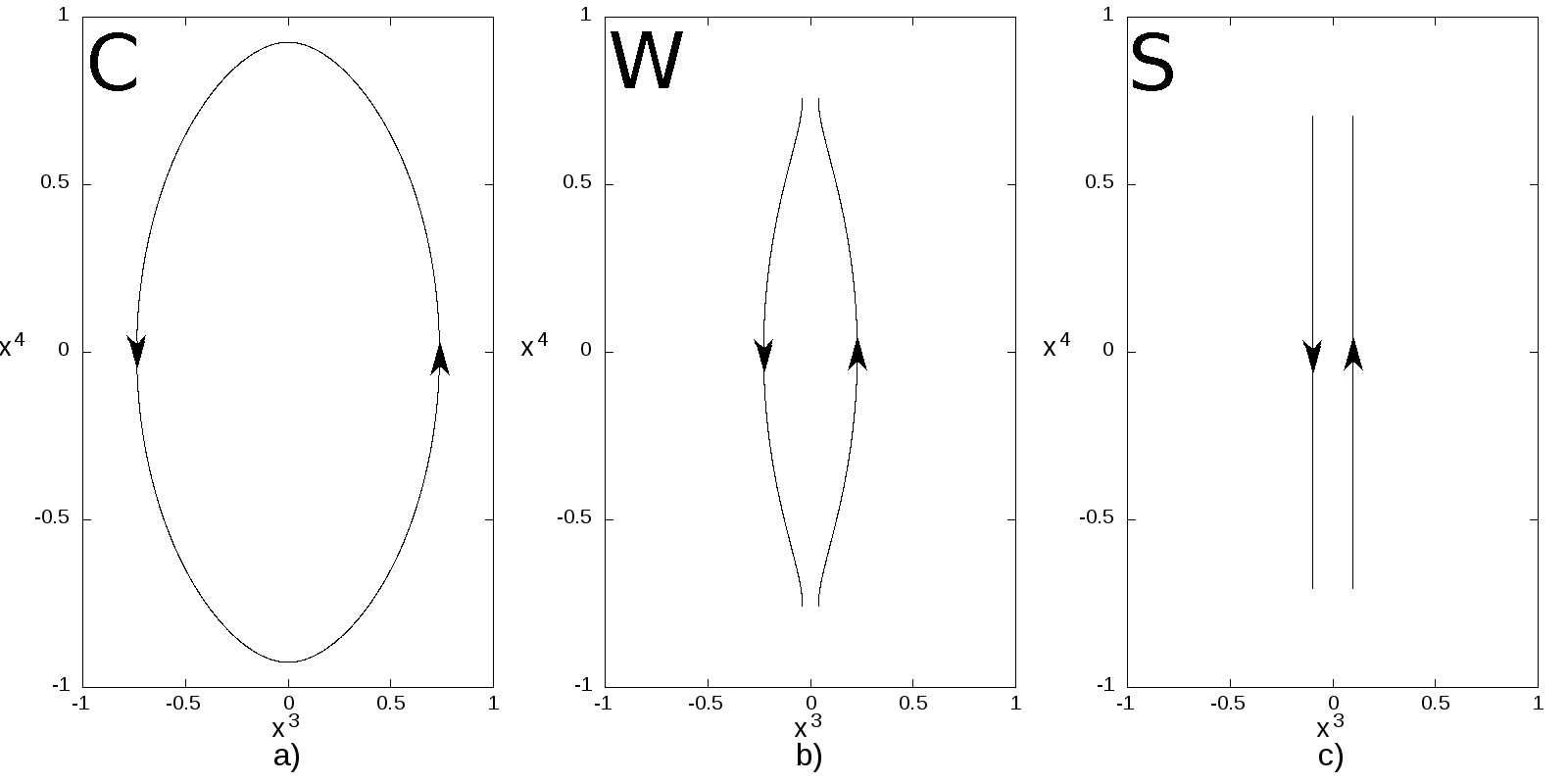}
  \caption{Examples of the three types of numerical solutions, all with $\kappa=0.5$ and $a=0.02$. From left to right: a) a C instanton with $\Temp=0.5$; b) a W instanton with $\Temp=0.66$; c) an S instanton with $\Temp=0.71\approx\Temp_{\lambda_1=0}$ (lower branch).}
  \label{fig:example_instantons}
\end{figure}

As discussed in section \ref{sec:regularisation}, we must regularise the interaction potential. This introduces a third parameter, $a$, the short distance cut off. For each point in the $(\kappa,\Temp)$ plane, we find the corresponding solution for a range of $a$ and evaluate the action, $\Action(\kappa,\Temp;a)$, and its Legendre transform, $\Entropy(\kappa,\Temp;a)$. For small enough $a$ we should be able to fit these to a linear function
\begin{equation}
\Action(\kappa,\Temp;a) \approx \Action(\kappa,\Temp) +  c(\kappa,\Temp) a, \label{eq:extrapolation}
\end{equation}
for some $c(\kappa,\Temp)$. To find $\Action(\kappa,\Temp)$ we then extrapolate to $a\to 0$, ensuring that the $a$ dependence is linear (see Fig. \ref{fig:aExtrapolation}).

The Newton-Raphson method does not converge in the presence of zero modes, essentially because the solution is not unique. As described in section \ref{sec:instantons}, we fix the translation zero modes by demanding that the centre of mass of the worldline is at the origin, $\bar{x}^\mu=0$. There is also a fifth zero mode corresponding to the remaining global symmetry of reparameterisation invariance. For worldlines with circular topology we fix this by demanding that $x^3_0-x^3_{N/2 - 1}=0$. Given a suitable initial guess this essentially fixes the point $x_0$ to be at the bottom of the loop and $x_{N/2 -1}$ to be at the top. In the high temperature case, where the instanton splits up into two separate worldlines, the global part of the reparameterisation invariance must be fixed on each side separately. We do so by demanding that there is a turning point at $x_0$ on the right hand side and at $x_{N-1}$ on the left hand side, i.e. we fix the spatial derivative to be zero there. In all cases we use Lagrange 
multipliers to impose constraints.

In the high temperature case we found that there is also a quasi-zero mode associated with translating one of the halves forward in Euclidean time and the other downwards. The presence of this quasi-zero mode slows the convergence of the Newton-Raphson method. To prevent this slowing down we fixed $\bar{x}_L^4=0$ and $\bar{x}_R^4=0$, rather than simply $(\bar{x}_L^4+\bar{x}_R^4)=0$ (subscripts $L$ and $R$ refer to left and right hand sides). This over-constrains the problem but the solutions thereby found are also solutions of the original problem. Further, from the parity symmetry we expect solutions to satisfy this extra constraint.

\begin{figure}
 \centering
  \includegraphics[width=0.48\textwidth]{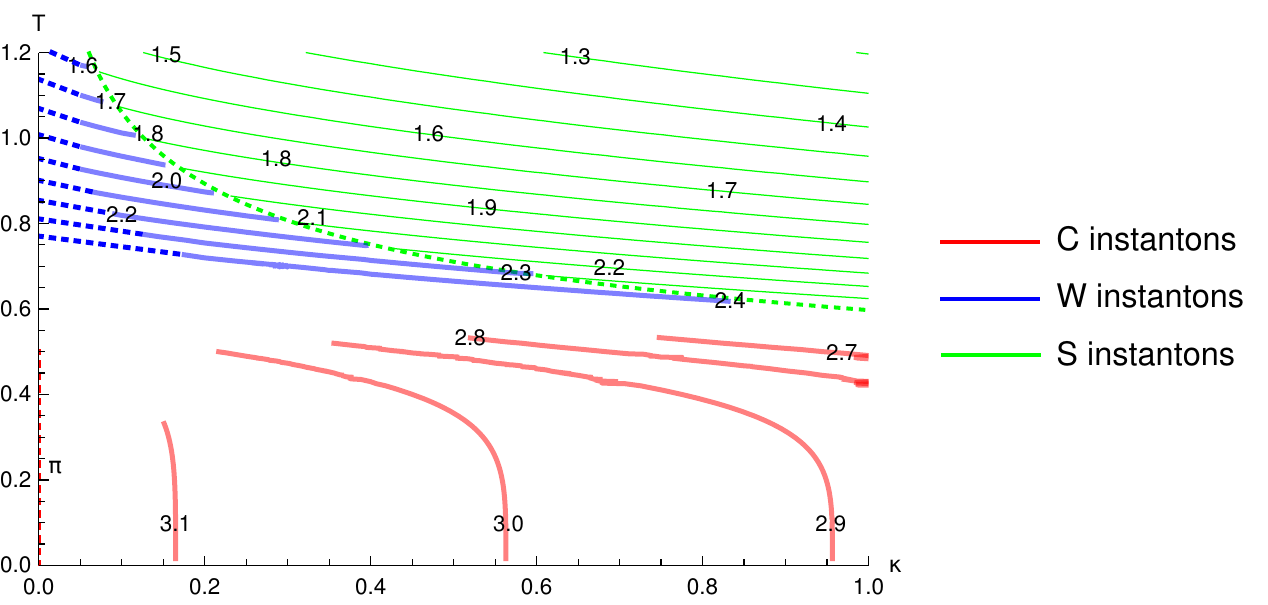}
  \caption{Contour plot of the action, $\Action(\kappa,\Temp)$, as calculated numerically. The solid red and blue lines are our numerical results and the solid green lines are given by Eq. \eqref{eq:action_straight}. The region in the top right, bounded by the green dotted line, is the region within which the S instanton is the only known solution. The blank region between the solid red and blue lines and for small $\kappa$ is where we could not maintain the hierarchies of Eq. \eqref{eq:scale_hierarchy} with $N=2^{12}$ points. The dashed blue lines are linear extrapolations from the contours found numerically to the same value of the action at $\kappa=0$ (Eq. \eqref{eq:s_lemon_kappa0}).}
  \label{fig:contour_S}
\end{figure}

In this way we can start to fill in the $(\kappa,\Temp)$ plane with instanton solutions, building up a contour plot of the action and a phase diagram. Each of the three different classes of solutions (C, W and S instantons) has a region of existence and a region within which it has the lowest Euclidean action (the actions denoted respectively by $\Action_C$, $\Action_W$ and $\Action_S$). If two solutions exist at a given point in the plane, that with lower Euclidean action determines the rate, and hence defines the phase. Only the regions with positive action can describe tunnelling processes. Fig. \ref{fig:contour_S} is a contour plot of the Euclidean action as calculated numerically. 

The phase diagram that emerges is quite interesting. The S instantons exist over the whole $(\kappa,\Temp)$ plane. The C and W instantons do not. Where we have found the W instantons to exist, they have lower action than the S instantons. It also seems to be the case for the C instantons. It is the case at $\kappa=0$ and we can give an argument for it at $\kappa=4\pi$. The action of the C instantons goes to zero at $(4\pi,0)$ whereas that of the S instantons goes to zero at $(4\pi,1/\pi)$. Further, if we can assume that $\Action_C$  and $\Action_S$ decrease with increasing temperature (i.e. the solutions have positive energy), then, where the C instantons exist for $\kappa=4\pi$ and $\Temp>0$, they must have lower action. We have not found numerically a region within which both the C and W instantons exist. It may be that they exist in disjoint regions, or it may be that they coexist near their phase boundary where our numerical calculations fail.

From Fig. \ref{fig:contour_S} we can see the existence of two lines of phase transitions: $\Temp_{CW}(\kappa)$ separating the C instantons from the W instantons and $\Temp_{WS}(\kappa)$ separating the W instantons from the S instantons. From our numerical results, within the range of parameters explored, the line defined by $\Temp=\Temp_{WS}(\kappa)$ appears to coincide with the lower branch of $\Temp_{\lambda_1=0}(\kappa)$. This line is a line of second order phase transitions, as discussed in section \ref{sec:hightemp}. The order of the phase transitions at $\Temp=\Temp_{CW}(\kappa)$ is not clear, except at $\kappa=0$ where it is of second order, as discussed in section \ref{sec:finite_coupling_results}. At $\kappa=0$ we also have that $\Temp_{CW}(0)=1/2$. Above this we can say nothing precise as, in the region around $\Temp=\Temp_{CW}$, we have not been able maintain the hierarchies of Eq. \eqref{eq:scale_hierarchy}. However it appears that $\Temp_{CW}(\kappa)\approx 1/2$, at least for $\kappa\leq 1$.

For $(\kappa,\Temp)$ outside the region spanned by our numerical calculations (see Fig. \ref{fig:contour_S}), there is little we can say about the form of the phase diagram. The two lines of phase transitions may cross at some point $\Temp_{CW}(\kappa_*)=\Temp_{WS}(\kappa_*)$, which we denote by $(\kappa_*,\Temp_*)$, or even at multiple points. Alternatively the line of phase transitions between C and W instantons may remain forever below that of W and S instantons, i.e. $\Temp_{CW}(\kappa)<\Temp_{WS}(\kappa)$ for all $\kappa$. More work is needed to better understand the phase diagram for larger $\kappa$ and $\Temp$.

\begin{figure}
 \centering
  \includegraphics[width=0.48\textwidth]{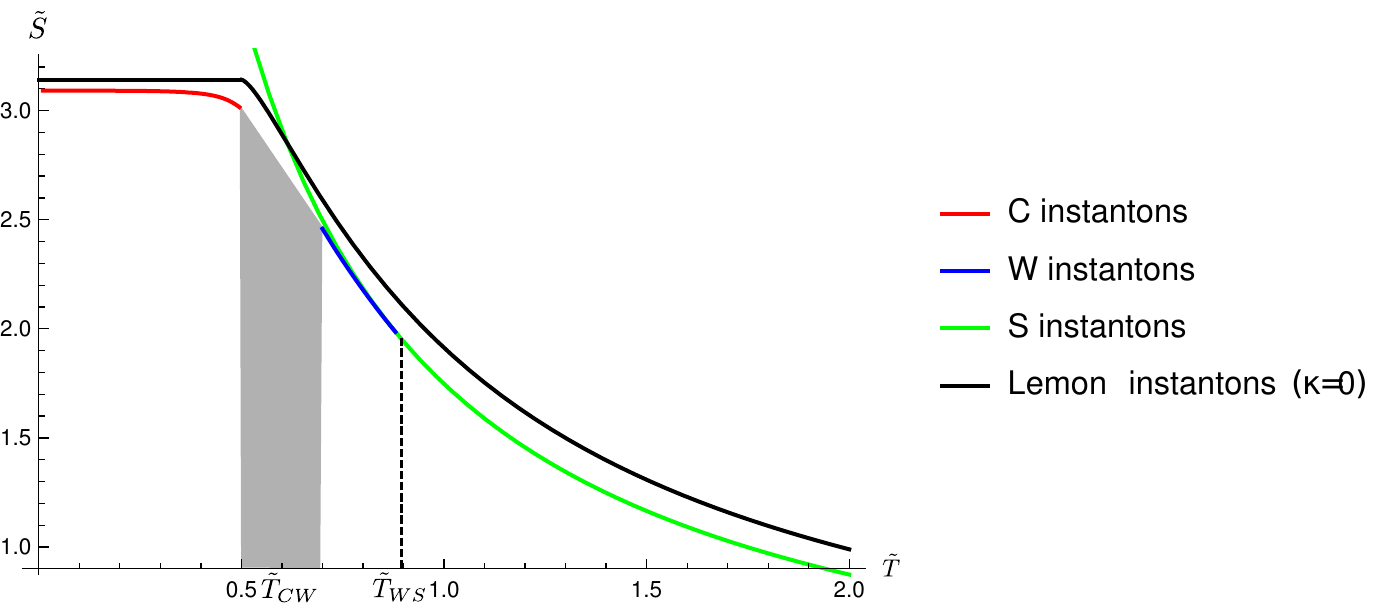}
  \caption{A slice through the $(\kappa,\Temp)$ plane at fixed $\kappa=0.2$. The action of the W instantons (blue) is lower than that of the S instantons (green) where they exist. The lemon shaped instantons are only exact solutions for $\kappa=0$. The expression for their action is $\pi$ for $\Temp<1/2$ and equation $\eqref{eq:s_lemon_kappa0}$ for $\Temp>1/2$. The transition temperature, $\Temp_{CW}$, lies somewhere in the grey shaded region. The difficulty in maintaining the hierarchies of Eq. \eqref{eq:scale_hierarchy} have prevented us from calculating it more accurately.}
  \label{fig:S_kappa_0.2}
\end{figure}

For comparison with the analytic results enumerated in sections \ref{sec:finitecoupling}, \ref{sec:lowtemp} and \ref{sec:hightemp}, in Fig. \ref{fig:S_kappa_0.2} we also give a plot comparing the action as a function of $\Temp$, for fixed $\kappa=0.2$.

\subsection{Fixed energy results}

\begin{figure}
 \centering
  \includegraphics[width=0.48\textwidth]{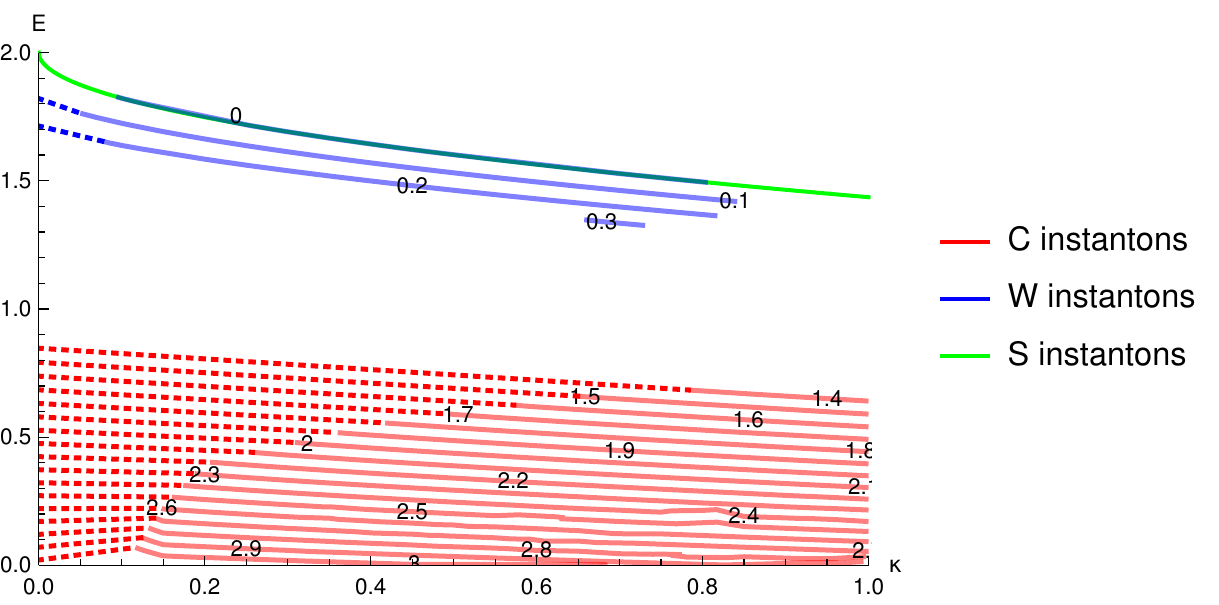}
  \caption{Contour plot of $\Entropy(\kappa,\Erg)$, the exponential suppression of the inclusive rate of Schwinger pair production at fixed energy. The blank region in the top right, bounded by the green line, is the region within which the exponential suppression of the S instanton is negative. The blank region between the solid red and blue lines and for small $\kappa$ is where we could not maintain the hierarchies of Eq. \eqref{eq:scale_hierarchy} with $N=2^{12}$ points. The dashed blue lines are linear extrapolations from the contours found numerically to the same value of $\Entropy$ at $\kappa=0$ (Eq. \eqref{eq:entropy_zero_kappa}).}
  \label{fig:contour_Sigma}
\end{figure}

We also calculate the Legendre transform of these results. To calculate the energy of a solution we use Eq. \eqref{eq:erg}. Fig. \ref{fig:contour_Sigma} is a contour plot in the $(\kappa,\Erg)$ plane of the exponential suppression, $\Entropy$.

At $\kappa=0$ the relevant instanton is the lemon shaped one discussed in section \ref{sec:finite_coupling_results}. The corresponding suppression is given by Eq. \eqref{eq:entropy_zero_kappa} and there is no phase transition for any $0<\Erg<\Erg_c$. On the other hand for $\kappa>0$ there is a phase transition between the C and W instantons.

In Fig. \ref{fig:contour_Sigma} we have also plotted extrapolations from our numerical results to the same value of $\Entropy$ at $\kappa=0$ (Eq. \eqref{eq:entropy_zero_kappa}). The extrapolations for both C and W instantons look good. How these instantons match onto the lemon instantons at $\kappa=0$, and where the phase transition between them lies, is not clear. Note that for small, non-zero $\kappa$ and small $\Erg$ the leading terms for both C and W instantons agree (equations \eqref{eq:entropy_small_kappa} and \eqref{eq:entropy_zero_kappa}).

\subsection{Numerical errors}

For a selection of our numerical solutions we performed various checks. For the C and W instantons we computed the lowest few eigenvalues of perturbations about the solutions and always found that there was one negative mode, as required for the solution to be interpreted as a tunnelling solution. We also computed the spectrum of eigenvalues about some S instantons, finding one negative eigenvalue for temperatures above the lower branch of $\Temp_{\lambda_1=0}$ and more than one below this temperature. The apparent absence of the self-force instability due to higher harmonic fluctuations may be due the cut-off, $a$, and due to the discretisation of the worldlines. The conservation of $\dot{x}^2$ was accurate to about 1 part in $10^4$ or better. The solutions were found to be symmetric under a rotation by $\pi$ in the 3-4 plane, to numerical accuracy.

The dominant errors in our numerical calculation are due to the difficulty of maintaining the hierarchies of Eq. \eqref{eq:scale_hierarchy}. We have rejected solutions for which $L[x]/(N a)>0.15$ or for which $a/\mathrm{Min}(\kappa,A^{-1}[x;i])>0.2$. The errors due to the finiteness of these quantities manifests in the extrapolation $a\to 0$ (see Eq. \eqref{eq:extrapolation}).

\begin{figure}
 \centering
  \includegraphics[width=0.4\textwidth]{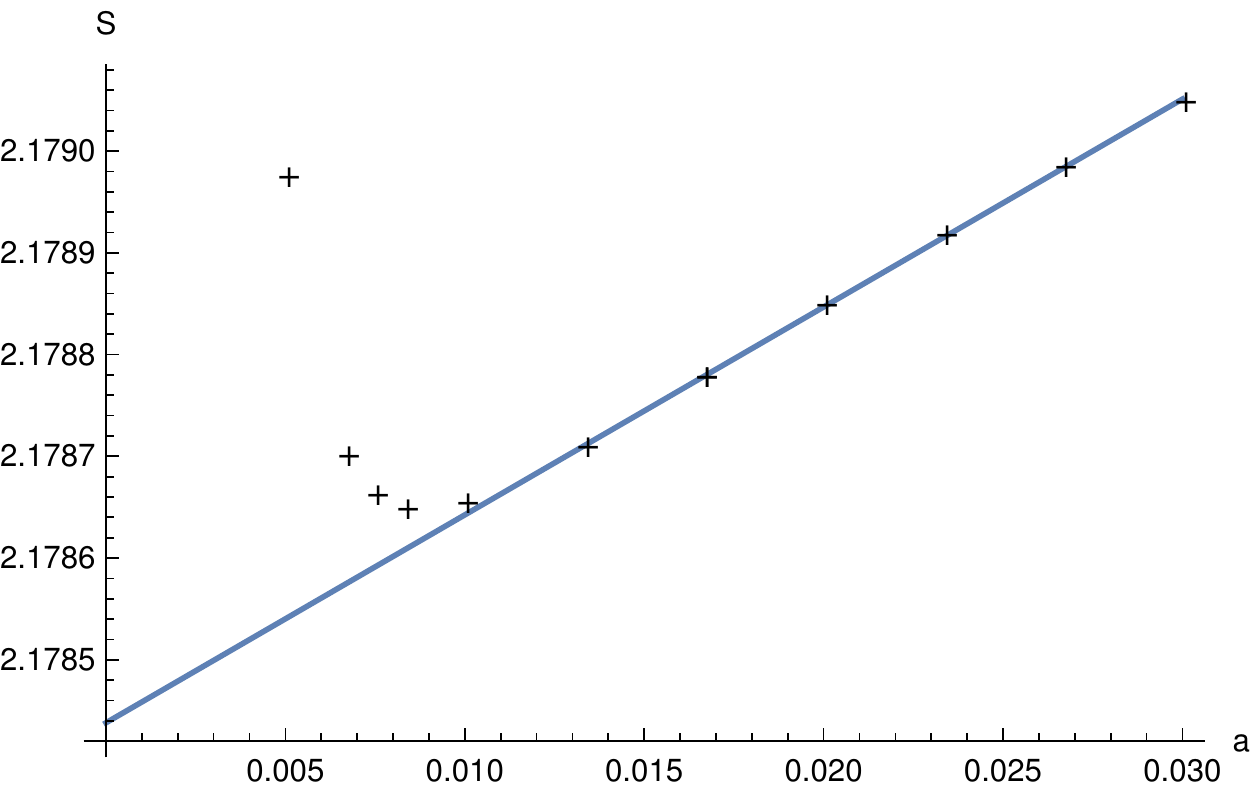}
  \caption{Plot of extrapolation of data to remove cut-off, $a\to 0$. Here are plotted data points for $\Action(\kappa,\Temp;a)$ for W instantons at $(\kappa,\Temp)=(0.2,0.8)$. The linear extrapolation is also plotted. The wild behaviour at very small $a$ is expected to be due to the effect of the discreteness scale, here $L[x]/N\approx 0.0006$.}
  \label{fig:aExtrapolation}
\end{figure}

See Fig. \ref{fig:aExtrapolation}. For very small $a$, the dependence of the action on $a$ is strongly nonlinear. This is due to $a$ becoming comparable with the distance between points, $L[x]/N$, and hence the discreteness of the representation of the worldline becomes significant. We implemented an algorithm to fit to only the linear part of the plot. For each point in the $(\kappa,\Temp)$ plane we assemble the data $\{a,\Action(\kappa,\Temp;a)\}$ in an array, ordered by the value of $a$. We then fit straight lines to all subsets of at least four consecutive data points, ensuring that this covers a range of $a$ such that the maximum value is at least twice the minimum value. For each fit we calculate the standard error in the result. For our final result, $\Action(\kappa,\Temp)$, we take that with least standard error. We also throw away results for which the standard error due to the fit is greater than $0.01$, though in most cases it is much smaller.

For $\Temp=0$ we have both approximate, numerical results and an exact, analytic expression, Eq. \eqref{eq:s_zerotemp}. The difference between the two is found to increase with $\kappa$ up to about $0.01$ at $\kappa=1$, using $N=2^{12}$ points. This error scales with the number of points as $1/N^2$. We also have an exact, analytic expression for large temperatures, Eq. \eqref{eq:action_straight}. Unfortunately though, for the corresponding instanton, the S instanton, due to the enhanced symmetry the zero modes corresponding to time translation and to reparameterisation invariance cannot be fixed as for the other instantons, preventing convergence of the Newton-Raphson method.

As we approach the phase transition between C and W instantons, $\Temp_{CW}(\kappa)$, it becomes more difficult to maintain the hierarchies of Eq. \eqref{eq:scale_hierarchy}. Hence we expect errors there to be greater.

\section{Conclusions} \label{sec:conclusions}

In this paper, we have extended previous results on Schwinger pair production to arbitrary couplings and arbitrary temperatures. To achieve this we restricted ourselves to weak, constant external fields. This restriction was shown to result in a semiclassical approximation and within this approximation we have calculated the leading behaviour of the logarithm of the rate. As a by-product, we were also able to obtain inclusive pair production rates at fixed energy.

We adopted the worldline description. In this framework the problem reduced to one of solving the instanton equations of motion for a self-interacting worldline, an interesting geometric problem.

For weak couplings, like in QED, our results complement the extensive literature on the subject, providing an alternative approach which holds at all temperatures and in which some issues are clearer. In this case $\kappa<\epsilon$ and our approach, which includes all orders in $\kappa$ but just the leading order in $\epsilon$, does not seem necessary. However, as we have discussed, the singular nature of the small $\kappa$ perturbation means that one aught not to simply set $\kappa=0$ from the outset. Doing this may lead to the incorrect instanton and hence to incorrect results at leading order in $\epsilon$.

In this weak coupling regime, and at temperatures $\Temp<1/2$, our results give small corrections to the leading order results. When expanded they capture the two-loop, thermal correction for weak fields. We also find no thermal correction at one-loop in this temperature range. Though we have calculated analytically the correction to the exponent to leading and next to leading order in $\kappa$, a full calculation of higher order loop corrections would require also the thermal corrections to the prefactor.

At higher temperatures, $\Temp>1/2$, the singular nature of the weak coupling expansion gives non-trivial corrections to the naive $\kappa\to 0$ limit. Our numerical solutions at finite $\kappa$ appear to approach the well known lemon shaped instanton as $\kappa\to 0$. At leading order, this solution gives a non-zero thermal enhancement to the rate. Note however that this enhancement is not present in the one-loop approximation, which breaks down due to the singular nature of the weak coupling expansion. The lemon shaped instanton also shows an enhanced correction to the exponential suppression of order $\kappa \log(\kappa)/\epsilon$. This dominates over order the $\epsilon^0$ correction for sufficiently small $\epsilon$.

At intermediate and strong couplings our results open new avenues. Using them one can make reliable estimates for the pair production rate of strongly charged particles via the thermal Schwinger process. In particular one could apply these results to the pair production of magnetic monopoles. Sufficiently light magnetic monopoles would be produced amply in the strong magnetic fields and high temperatures present in heavy ion collisions, in neutron stars and in the early universe.

In this paper we have only calculated the exponential suppression of the rate. For direct phenomenological application, one should also calculate the prefactor. The Refs. \cite{ivlev1987tunneling,garriga1994instantons,gordon2014worldline} would be an apt place to start. They all find similar instantons to ours, though in theories without dynamical long-range forces.

The appearance of the self-force instability in the semiclassical evaluation of the path integral raises some intriguing questions that require further work. So too does the instanton phase diagram, Fig. \ref{fig:contour_S}, for which further work is needed to determine the form of the phase diagram at larger values of $\kappa$ and $\Temp$.

The worldline description that we have developed here could be used to calculate pair production rates for other induced Schwinger processes at arbitrary coupling. For example one could consider a non-constant external field. The numerical approach we have adopted would then directly apply.

\begin{acknowledgments}
The authors would like to thank Toby Wiseman, Sergey Sibiryakov, Ian Jubb, Edward Gillman, and Lois Overvoorde for illuminating discussions. OG would also like to thank Lois Overvoorde for help with the computational work. OG is supported by an STFC studentship and AR by STFC grant ST/L00044X/1. The computational work was undertaken on the Cambridge COSMOS SMP system, part of the STFC DiRAC HPC Facility supported by BIS NeI capital grant ST/J005673/1 and STFC grants ST/H008586/1, ST/K00333X/1.
\end{acknowledgments}

\appendix
 
\section{QED and SQED in weak external fields\label{appendix:qedsqed}} 

For sufficiently small scalar self-coupling, $\lambda$, the only difference between QED and SQED is the spin of the charged particles. Following the manipulations of SQED in Eq. \eqref{eq:pathintegral1} we see that the difference for a spin $1/2$ charged particle manifests simply in a replacing the trace over the Klein-Gordon operator with a trace over the Dirac operator. As in Refs. \cite{feynman1951operator,polyakov1988two}, we note that this difference can be taken into account by including the spin-factor in the path integral representation of the functional trace,
\begin{align}
 \mathrm{Tr}& (\mathrm{e}^{-(-i\slashed{D}-i m)s}) = \nonumber \\
 &\int \mathcal{D}x^\mu \mathrm{e}^{-S_0[x^\mu,A^{ext}_\mu+A_\mu;s]}Spin[x^\mu,A^{ext}_\mu+A_\mu;s],
\end{align}
where $S_0[x^\mu,A^{ext}_\mu+A_\mu;s]$ is given in Eq. \eqref{eq:s0_definition} and the spin factor is given by
\begin{equation}
 Spin[x^\mu,A^{ext}_\mu+A_\mu;s]:= Tr_\gamma \mathcal{P} \ \mathrm{e}^{ i g \int_0^s \mathrm{d}\tau \Sigma^{\mu\nu} (F^{ext}_{\mu\nu}(x)+F_{\mu\nu}(x))}
\end{equation}
where $ Tr_\gamma $ signifies the trace over spinorial indices, $\mathcal{P}$ is the path ordering operator and $\Sigma^{\mu\nu}$ are the generators of Lorentz transformations in the spin $1/2$ representation, i.e. $\Sigma^{\mu\nu}=[\gamma^\mu,\gamma^\nu]/2$, where $\gamma^\mu$ are the gamma matrices. The next step is to integrate over the gauge field $A_\mu$. Note that even with the inclusion of the spin factor the integration over $A_\mu$ is Gaussian and hence can be done exactly. In the spin $0$ case, the integration takes the following form
\begin{equation}
\int \mathcal{D}A_\mu \ \mathrm{e}^{ -\frac{1}{2}\int_x\int_y A^\mu(x)G^{-1}_{\mu\nu}(x,y)A^\nu(y) + i \int_x A_\mu(x)j^\mu_0(x)},
\end{equation}
where $\int_x:=\int \mathrm{d}^4x$, and $j^\mu_0(x)$ is given by
\begin{equation}
j^\mu_0(x) = g \int_0^s\mathrm{d}\tau \dot{x}^\mu (\tau) \delta^{(4)}(x-x(\tau)).
\end{equation}
Performing the integration leads to the following exponential
\begin{equation}
\exp\bigg\{ -\frac{1}{2}\int_x \int_y j^\mu_0(x)G_{\mu\nu}(x,y)j^\nu_0(y)\bigg\}.
\end{equation}
The difference in the spin $1/2$ case amounts to the replacement
\begin{align}
j^\mu_0(x) &\to j^\mu_0(x)+i g \int_0^s\mathrm{d}\tau \Sigma^{\mu\nu}\partial_\nu \delta^{(4)}(x-x(\tau)), \nonumber \\
& \ \ \ := j^\mu_0(x) + \xi^\mu(x).
\end{align}
Now, we scale all the dimensionful quantities as in section \ref{sec:general}, i.e. $\tau \to \tau/s$, $s\to s/gE$ and $x^\mu \to x^\mu m/gE$. This reduces all dependence on the parameters to dependence on $\epsilon := gE/m^2 \ll 1$ and $\kappa := g^3E/m^2$. We can now write the interaction terms in the spin $1/2$ case as
\begin{align}
\exp\bigg\{-\frac{\kappa}{2\epsilon}&\int_x \int_y j^\mu_0(x)G_{\mu\nu}(x,y)j^\nu_0(y) \nonumber \\
&- \kappa\int_x \int_y \xi^\mu(x)G_{\mu\nu}(x,y)j^\nu_0(y) \nonumber \\
&- \frac{\kappa\epsilon}{2}\int_x \int_y \xi^\mu(x)G_{\mu\nu}(x,y)\xi^\nu(y)\bigg\},
\end{align}
where $j^\mu$, $\xi^\mu$ and $G_{\mu\nu}$ are now independent of $g$, $E$ and $m$. In this paper, we have allowed $\kappa$ to freely vary up to $O(1)$ (that is because we only require $\kappa \ll g^2$ and we consider strong coupling) however for the semiclassical approximation to be valid we require $\epsilon \ll 1$. Hence the spin dependent factors are subleading (as long as the dimensionless parts are at most $O(1)$) and we can drop all $\xi^\mu$ dependence.

The net result of all this is that, to leading order in $\epsilon$, the instanton describing pair production is the same for both theories, as is the fluctuation prefactor about the instanton, excluding an overall spin factor $(2s+1)$. Charge renormalisation, which is not included to leading order in $\epsilon$, is different in QED and SQED. In both cases we expect the final results to depend on the renormalised charge, as discussed in section \ref{sec:regularisation}. 
 
\section{Worldline description of extended particles\label{appendix:extended}} 

For elementary particles the geometric worldline description arises naturally and can be derived by standard methods from the field theoretic description, as shown in section \ref{sec:general}. For extended field configurations, such as solitons, no exact worldline description can exist. However, for circumstances where the extended field configuration is much smaller than all other scales, an effective worldline description can suffice \cite{selivanov1985destruction,voloshin1991illustrative,affleck1981monopole}. This is analogous to the description of the motion of planets in the solar system in terms of the motion of points.

In \cite{affleck1981monopole} just such an effective description was explicitly derived for the 't Hoof-Polyakov monopole \cite{polyakov1974particle,thooft1974magnetic}. The worldline instanton that they found was a circle and the effective worldline description was found to be valid when the radius of the circle was much larger than the size of the 't Hooft-Polyakov monopole. We wish to generalise this result for more general worldline curves.

The 't Hooft-Polyakov monopole is a static solution to the field equations for the Georgi-Glashow model, and other similar theories. It is an extended solution and hence it cannot be said to be \emph{at} a position, however it does have a well defined centre and core region, beyond which all but the Abelian gauge field is exponentially damped. We can thus assign to the centre of the monopole solution a worldline, i.e. a map from the real line to the path in Minkowski space traced out by the centre of the monopole solution. The static solution and Lorentz transformations of it (which are also solutions) have straight, timelike worldlines. Static solutions of the Euclidean (Wick rotated) theory need not be timelike.

Combinations of straight worldlines are no longer exact solutions due to the interactions between them. However, in the limit that the monopoles are infinitely separated this should become an exact solution \cite{bogomolny1980calculation} \footnote{In the Bogomolny-Prasad-Sommerfeld limit \cite{prasad1975exact}, i.e. when the scalar self-coupling is infinitesimally small and positive, such superpositions of static, like-charged monopoles are in fact exact solutions. That is because in this limit the attraction due to the Higgs field is exactly cancelled by the repulsion due to the gauge field \cite{manton1977force}. At low energies the dynamics of such multi-monopole solutions is given by geodesic motion of the collective coordinates on the configuration space of solutions \cite{manton1981remark,atiyah1985low,gibbons1986classical}.}. At finite separation, due to the long ranged interaction between monopoles, the solution is no longer exact. However, we can find an approximate solution following \cite{
affleck1981monopole}.

We consider for example the Georgi-Glashow theory. To find an approximate solution to the full (Euclidean) field equations we first solve the equations of motion for a pointlike monopole in a given external magnetic field, at a certain temperature, including the self-interaction. These are the classical worldline calculations we have carried out in this paper. For simplicity, we restrict the worldline to the $(x_3,x_4)$ plane. We use construct coordinates centred on the worldline and Fermi-Walker transported along it (see for example \cite{misner1973gravitation}). We denote the coordinate along the worldline as $u$ and the normal coordinate in the plane as $v$, with $(x_1,x_2,v)=(0,0,0)$ on the worldline. Other than photon excitations, which are taken into account, internal excitations of the 't Hooft-Polyakov monopole are gapped and hence, for sufficiently low energies, we can assume these are not excited. That is, we can assume translation invariance along the monopole worldline. In this case, the field 
equations near the 
monopole worldline read
\begin{align}
 D_iF^{ij}+a(u)F^{vj}+O(a(u)^2v^2) &= [D_i\phi,\phi], \nonumber  \\
 D_iD^i\phi+a(u)D_v\phi+O(a(u)^2v^2) &= \frac{\lambda}{g^2}(|\phi|^2-M_W^2)\phi,
\end{align}
where $D_a = \partial_a+igA_a$; $i,j$ run over $(x_1,x_2,v)$; $\lambda$ is the four point self-coupling of the Higgs particle; $M_W$ is the mass of a W boson and $a(u)$ is the magnitude of the acceleration of the worldline. At zeroth order in the acceleration these equations are solved by the 't Hooft-Polyakov magnetic monopole, static along the worldline. Hence, at lowest order in the acceleration the full field theoretic calculation reduces to the geometric, worldline one which we have pursued in this paper. This also requires that the radius of curvature of the worldline is large compared with the classical radius of the monopole solution. In our dimensionless units this is $\kappa$ (for the Georgi-Glashow theory) and hence we get the constraint $\kappa\ll a(u)^{-1}$.
 
\section{Finite difference formulation\label{appendix:numerics}} 

In this appendix we give our discrete approximation to the action in Eq. \eqref{eq:geometric_action} and the corresponding equations of motion. We use a simple finite difference approximation
\begin{align}
\Action[x]&=\sqrt{N\sum_i (x^\mu_{i+1}-x^\mu_i)^2} - \sum_i x^3_i (x^4_{i+1}-x^4_i)\nonumber \\
&-\frac{\kappa}{2} \sum_{i,j}(x^\mu_{i+1}-x^\mu_i)(x^\mu_{j+1}-x^\mu_j)G_R(x_i,x_j;\Temp,a) \label{eq:numerical_action}
\end{align}
where $i$ and $j$ run over $0,1,...,N-1$ and contractions of Lorentz indices are implied. As discussed in section \ref{sec:regularisation} we choose an exponential counterterm, rather than the simpler length counterterm of Polyakov, so that the bare mass is positive. Summing the infinite periodic copies of the regularised propagator gives
\begin{align}
\sum_{n=-\infty}^{\infty} &\frac{-1}{4\pi^2((t+n/\Temp)^2+r^2+a^2))} = \nonumber \\
& \frac{\Temp\sinh \left(2 \pi \Temp \sqrt{a^2+r^2} \right)}{4 \pi    \sqrt{a^2+r^2} \left(\cos
   \left(2 \pi \Temp t \right)-\cosh \left(2 \pi \Temp  \sqrt{a^2+r^2} \right)\right)},
\end{align}
and likewise for the exponential counterterm
\begin{align}
\sum_{n=-\infty}^{\infty} \frac{\sqrt{\pi}}{4\pi^2a^2}&\mathrm{e}^{-(r^2+(t+n/\Temp)^2)/a^2}= \nonumber \\
&\frac{\Temp \mathrm{e}^{-r^2/a^2} \vartheta _3\left(\pi \Temp t,\mathrm{e}^{- \pi ^2 a^2 \Temp^2}\right)}{4 \pi  a }
\end{align}
where $t$ and $r$ are the temporal and spatial differences as in section \ref{sec:finite_temp_rate} and $\vartheta_3$ is the Jacobi theta function of the third kind. Due to the lack of well optimised numerical libraries for the Jacobi theta function, we in fact make a different choice of counterterm, which also reduces to the exponential regularisation for small $\Temp$,
\begin{align}
&G_R(x_i,x_j;\Temp,a) = \nonumber \\
&\frac{\Temp\sinh \left(2 \pi \Temp \sqrt{a^2+r^2} \right)}{4 \pi   \sqrt{a^2+r^2} \left(\cos
   \left(2 \pi \Temp t \right)-\cosh \left(2 \pi \Temp \sqrt{a^2+r^2} \right)\right)} \nonumber \\
   &\qquad\qquad\ +\frac{\sqrt{\pi} \mathrm{e}^{-r^2/a^2}\mathrm{e}^{\left(\cos \left(2 \pi \Temp t \right)-1\right)/(2 \pi ^2 a^2 \Temp^2)}}{4\pi^2 a^2 }
&
\end{align}
This regularisation is smooth and periodic in $1/\Temp$, as well as being relatively fast to numerically evaluate.

One further point, also mentioned in section \ref{sec:regularisation}, is that there is no need to regularise the interactions between disconnected parts of worldlines, one may use the unregularised propagator. This is useful as it removes some sources of error due to the finite cut off, $a$. In our calculations, we have used the unregularised propagator for the interaction between the left and right hand sides of the W instantons. It could also be used, though we didn't, for the interaction between thermal copies for the C instantons.

As discussed in section \ref{sec:numerical_results}, we fix the $N_0$ zero and quasi zero modes using Lagrange multipliers. Writing the constraint equations as $C_a[x]=0$, where $a$ runs over $1,..,N_0$, we define a new action including the Lagrange multiplier terms
\begin{equation}
 \Action[x,\lambda]:=\Action[x]+\sum_{a=1}^{N_0}\lambda_a C_a[x]. \label{eq:numerical_action_lagrange}
\end{equation}
The $\lambda_a$ are the Lagrange multipliers.

The equations of motion, which are simply $4N+N_0$ coupled, nonlinear, algebraic equations are found by taking partial derivatives of \eqref{eq:numerical_action_lagrange}, with respect to $x^\rho_k$ and $\lambda_b$, 
\begin{align}
\frac{\partial \Action[x]}{\partial x^\rho_k}&=0, \nonumber \\
C_b[x]&=0.
\end{align}
The Newton-Raphson equations derived from these, and an initial guess, are a system of linear equations, which we solve by LU decomposition, using the numerical library Eigen 3 \cite{eigen}. The 1 and 2 directions are trivial and decouple, leaving $2N+\tilde{N}_0$ equations, where $\tilde{N}_0(<N_0)$ is the number of zero modes in the reduced space.

\section{Numerical data\label{appendix:numerical_data}} 

Along with this paper, we have made available the numerical results presented in summary form in section \ref{sec:numerical_results}. They are in the file \emph{gould2017thermal\_results.csv}. The first line gives the column headings and all following lines give the corresponding numerical data as comma separated values. The meanings of the headings are as follows.

\begin{itemize}
\setlength\itemsep{-0.3em}
\item[] pot: The nature of the interaction potential, see below.
\item[] log2N: $log_2(N)$, where $N$ is the  number of points in the worldline.
\item[] kappa: the coupling, $\kappa:=g^3E/m^2$.
\item[] T: the temperature, $\Temp:=mT/gE$.
\item[] a: the cut-off, $a$.
\item[] S: the action, $\Action:=S/\epsilon$.
\item[] E: the energy, $\Erg:=\mathcal{E}/m$.
\item[] len: the length of the worldline.
\item[] kinetic: the gauge-fixed length of the worldline, $\tilde{L}$, Eq. \eqref{eq:gaugefixedlength}.
\item[] i0: minus the area of the worldine, $-\tilde{A}$, Eq. \eqref{eq:area}.
\item[] vr: the interaction, $\tilde{V}_R$, Eq. \eqref{eq:regularisation_rajantie} and the finite temperature generalisations.
\item[] zmax: The maximum distance between points in the $x^3$ direction.
\item[] zmin: The minimum distance between points in the $x^3$ direction.
\item[] tmax: The maximum distance between points in the $x^4$ direction. 
\item[] sol: The ratio of the norms of $\partial \Action[x]/\partial x^\rho_k$ and $x^\rho_k$.
\item[] acc\_max: The maximum acceleration along the worldline, see below.
\end{itemize}

The first column, pot, takes three different values depending on the nature of the interaction potential. It takes the value $1$ for zero temperature; $13$ for the low temperature, C instanton topology and $15$ for the high temperature, W instanton topology.

The acceleration referred to in the last column is a finite difference approximation to $|\ddot{x}|$, defined by
\begin{equation}
N^2|2x_k-x_{k+1}-x_{k-1}|,
\end{equation}
where $N$ is the number of points in the worldline.

\bibliography{thermalSchwinger9.bib}

\end{document}